\documentclass[a4paper,11pt]{article}
\usepackage{jheppub} % for details on the use of the package, please see the JINST-author-manual
\usepackage{lineno}
\usepackage{extarrows}
\usepackage{color}
\usepackage{helvet}         % selects\textbf{\textbf{•}} Helvetica as sans-serif font
\usepackage{courier}        % selects Courier as typewriter font
\usepackage{type1cm}        % activate if the above 3 fonts are
\usepackage{framed}
\usepackage{tikz}
\usepackage{makeidx}         % allows index generation
\usepackage{graphicx}        % standard LaTeX graphics tool
\usepackage{multicol}        % used for the two-column index
\usepackage[bottom]{footmisc}% places footnotes at page bottom

\usepackage{amsmath}
\usepackage{amssymb}
\usepackage{bbold}
\usepackage{amsthm}
\usepackage{subcaption}
\usepackage{sidecap}
\usepackage{floatrow}
\usepackage{pdflscape}
\usepackage{pdflscape}
\usepackage{comment}
\usepackage{enumitem}
\usepackage{esint}

\arxivnumber{2403.18576
} % if you have one

\title{Logarithmic correlation functions in 2D critical percolation}

\author[a,b]{Federico Camia}
\author[c]{Yu Feng}

\affiliation[a]{Science Division, New York University Abu Dhabi,\\
Saadiyat Island, Abu Dhabi, United Arab Emirates}
\affiliation[b]{Courant Institute of Mathematical Sciences, New York University,\\
251 Mercer Street, New York, NY 10012, U.S.A.
}
\affiliation[c]{Department of Mathematics,Tsinghua University,\\
Beijing 100084, People’s Republic of China}
\emailAdd{federico.camia@nyu.edu}
\emailAdd{yufeng\_proba@163.com}

\abstract{%The large scale geometric properties of two-dimensional critical percolation are believed to be described by a logarithmic conformal field theory, but finding examples of logarithmic singularities and a physical interpretation for their appearance in terms of lattice observables has been challenging.
	It is believed that the large-scale geometric properties of two-dimensional critical percolation are described by a logarithmic conformal field theory, but it has been challenging to exhibit concrete examples of logarithmic singularities and to find an explanation and a physical interpretation, in terms of lattice observables, for their appearance.
	We show that certain percolation correlation functions receive independent contributions from a large number of similar connectivity events happening at different scales.
	Combined with scale invariance, this leads to logarithmic divergences.
	We study several logarithmic correlation functions for critical percolation in the bulk and in the presence of a boundary, including the four-point function of the density (spin) field.
	Our analysis confirms previous findings, provides new explicit calculations and explains, in terms of lattice observables, the physical mechanism that leads to the logarithmic singularities we discover.
	Although we adopt conformal field theory (CFT) terminology to present our results, the core of our analysis relies on probabilistic arguments and recent rigorous results on the scaling limit of critical percolation and does not assume a priori the existence of a percolation CFT.
	As a consequence, our results provide strong support for the validity of a CFT description of critical percolation and a step in the direction of a mathematically rigorous formulation of a logarithmic CFT of two-dimensional critical percolation.}
	

\begin{document}
\maketitle
\flushbottom

\section{Introduction}

\subsection{Motivation and description of the main results}

It is strongly believed that the large-scale geometric properties of two-dimensional critical percolation are described by a conformal field theory (CFT). It is moreover accepted that the proper theory must be a logarithmic conformal field theory (LCFT), but direct evidence or consequences of this fact and an explanation in terms of lattice observables of the appearance of logarithmic singularities have been challenging to find (see, for instance, the discussion in the introductions of \cite{VJS12} and \cite{GV18}).

From an algebraic perspective, it is well understood that LCFTs are characterized by the fact that the dilatation operator $L_0$ is non-diagonalizable and has a Jordan cell structure (see, e.g.,~\cite{VJS11}).
From this perspective, LCFTs can be analyzed using a powerful algebraic approach which, roughly speaking, consists in studying the indecomposable modules of the Virasoro algebra.
In recent years, this approach, combined with numerical techniques and conformal bootstrap methods, has led to tremendous progress, both on the side of a general theory of LCFTs and for specific models~\cite{JS19,NRJ23,PRS19,HJS20,NR21}.
In particular, Nivesvivat, Ribault and Jacobsen have recently used this approach to propose an analytic form for a class of four-point functions in various two-dimensional critical loop models, including the $O(n)$ and Potts models \cite{NRJ23}.

When the algebraic approach and bootstrap techniques are used to study a specific lattice model, it is implicitly assumed that the continuum (scaling) limit of the model admits a CFT description, and certain additional assumptions are usually made on the spectrum of the putative CFT.
In practice, when explicit expressions for correlation functions are found, they are typically obtained by solving differential equations derived from the Ward identities.
This approach doesn't usually explain the physical mechanism leading to the appearance of logarithms in terms of the lattice variables and observables of the original model.
Indeed, despite the progress mentioned above, exhibiting concrete examples of correlation functions with logarithmic divergences, and explaining the origin of such singularities in terms of lattice observables, has been a challenge.

In the case of percolation, there is an additional difficulty, and one more layer of abstraction, since percolation cannot be treated directly with the methods mentioned above, but needs to be defined as the formal $Q \to 1$ limit of the $Q$-state Potts model.
This formal limit, while considered standard, can be very delicate and remarkably subtle (see, for example, the discussions in~\cite{CR13} and~\cite{Dot20}).
%Formally, percolation corresponds the limit $Q \to 1$ of the $Q$-state Potts model, but this limit is very delicate and remarkably subtle, as observed, for example, in \cite{CR13} and~\cite{Dot20}.
%Indeed, the percolation partition function is identically $1$ and the techniques that work for the Potts model for $Q \neq 1$ do not usually work for percolation.
%so one can attempt to calculated percolation correlation functions by taking the $Q \to 1$ limit of Potts correlation functions.
%While this limit is standard, it is also very delicate and remarkably subtle, as observed, for example, in \cite{CR13} and~\cite{Dot20}.}

In this paper, we analyze several examples of percolation correlation functions with a logarithmic singularity with the help of a new approach, which allows us to avoid algebraic considerations and the $Q \to 1$ limit\footnote{As a consequence, a large part of our analysis is amenable to a fully rigorous mathematical formulation, which is presented in~\cite{CF24}.} and reveals the physical mechanism behind the logarithmic singularities.
Our starting point resembles the geometric considerations in \cite{VJS12} and \cite{GV18} in that we express correlation functions in terms of cluster connectivities, but our analysis then differs significantly from previous work on percolation that has appeared in the physics literature.
%In particular, we do not make any a priori assumptions on the operator content of the percolation CFT and we do not use the $Q \to 1$ limit of the $Q$-state Potts model nor Coulomb gas techniques.\footnote{As a consequence, our analysis is amenable to a fully rigorous mathematical formulation, which will be presented in a forthcoming paper~\cite{CF24}.}
More precisely, we show that certain four-point functions receive independent contributions from similar connectivity events at different scales.
Combined with scale invariance, this mechanism leads to logarithmic divergences as two of the four points collide.
By using only percolation observables and techniques, we give direct evidence of the logarithmic nature of the percolation CFT both in the bulk and in the upper half-plane, while providing new insight and a geometric explanation for the presence of logarithmic terms in critical percolation.
%Our methods and results therefore complement and extend the existing literature.
%and provide support for some of the assumptions implicit in the CFT approach to the study of critical percolation.
%provides new insight and an analytic/geometric explanation for the presence of logarithmic terms in critical percolation.
%and show how probabilistic techniques can be a useful tool.
%Although we do not discuss it here, part of our analysis can be extended to other geometric models, such as the Fortuin-Kasteleyn (FK) random-cluster model, and to dimensions higher than two.
%The latter appears as a consequence of having $c=0$, which manifests itself in the independence of percolation, combined with scale invariance.

The first quantity we will consider is the four-point function of the percolation density (spin) field in the bulk, which can be expressed as a linear combination of connection probabilities, the most fundamental percolation quantities.
We will show that this four-point function has a logarithmic divergence, as two of the four points collide, and will argue that this singularity is present also in the operator product expansion (OPE) of two density fields, in such a way as to imply the presence, in the right-hand side of the OPE, of two new fields with the same scaling dimension.
This is consistent with the idea, proposed by Gurarie in \cite{Gur93}, that LCFTs are characterized by the presence of pairs of fields with the same scaling dimension.
Importantly, we do not have to assume that the spectrum of the theory contains these two fields; instead, their presence is implied by our asymptotic analysis of the four-point function.

The new fields appearing in the OPE of two density fields can be identified with the energy field (discussed, for instance, in Section~4.2 of~\cite{Car13}) and with a mixture of the energy field and the four-leg operator.
The structure of the two-point functions of the new fields shows that they form a logarithmic pair \cite{Gur93,CR13}.
In particular, we show that the two-point function of the energy field is zero and explain how this is related to its normalization and the presence of a logarithm in the four-point function and the OPE.

In our second example, we will consider the four-point function in the upper half-plane between four density fields on the boundary, as well as the OPE obtained from the collision of two density fields on the boundary.
Once again, the four-point function has a logarithmic divergence, as two of boundary points collide, and the OPE shows the appearance of two new fields.
In this case, one of the new fields appearing in the OPE can be identified with the boundary stress-energy tensor.
The fact that a second field with the same scaling dimension also appears provides a test for Gurarie's proposal that the existence of a logarithmic partner to the stress-energy tensor is necessary (at least in certain circumstances) to avoid the so-called ``$c=0$ catastrophe'' \cite{Gur93}.

Our last example is the four-point function between two boundary three-leg operators and two boundary one-leg operators.
We derive the asymptotic behavior as various pairs of points collide and show that in one of these limits, the four-point function has a subleading logarithmic divergence.
Interestingly, the logarithmic term can be written as the integral of a three-point function.

\subsection{Background and connections to other works}

Percolation provides the simplest example of a purely geometric phase transition.
%Critical percolation is the simplest example of a geometric critical model.
%Unlike in models such as Ising and Potts, defined in terms of a local spin field,
In percolation and other geometric models, such as the Fortuin-Kasteleyn (FK) random cluster model, the focus is on connectivity properties.
These are encoded in connection probabilities, which replace the spin correlation functions of the Ising and Potts models as fundamental quantities of interest.
At the critical point, the large-scale geometric properties of percolation are believed to be described by a CFT.
This belief is based on the assumption, recently confirmed \cite{Camia24}, that, in the continuum limit, connection probabilities behave like CFT correlation functions.

If one postulates a CFT description of the large-scale properties of critical percolation, then the relevant CFT must have central charge $c=0$ because the percolation partition function is not sensitive to finite size effects \cite{BCN86,AFF86}.
In order to be nontrivial, a $c=0$ CFT must be non-unitary, since the only unitary CFT with $c=0$ does not admit any observables other than the identity field.
%The percolation partition function is not sensitive to finite size effects, so the percolation CFT must have central charge $c=0$~\cite{BCN86,AFF86} and, in order to be nontrivial, it must be non-unitary, since the only unitary CFT with $c=0$ does not admit any observables other than the identity field.

Lack of unitarity has serious physical implications, it can complicate the mathematical analysis and lead to the appearance of logarithmic singularities in some correlation functions \cite{RS92,Sal92,Gur93} and in the OPE of certain fields \cite{GUr99,VA02}.
From a lattice perspective, this behavior of critical correlation functions is somewhat mysterious.
The exponential decay of correlations away from the critical point and the appearance of power laws at the critical point are natural and easy to understand. The first is linked to the presence of a characteristic length while the second is a manifestation of scale invariance, i.e., the lack of a characteristic length.
From this perspective, logarithmic terms at the critical point appear as an anomaly, and the approaches mentioned above (the algebraic approach and taking the $Q \to 1$ limit of Potts model correlation functions) have limited power in explaining the physical mechanism behind those terms.

%Logarithmic singularities in CFT correlation functions were observed in~\cite{RS92,Sal92}, but it was Gurarie \cite{Gur93}.
%The fact that the structure of general CFTs allows the presence of multiplicative logarithms in four-point correlation functions was first observed in~\cite{RS92,Sal92, Gur93}. 
%argued that it can lead to the appearance of logarithmic divergences in some four-point correlation functions .
%(see also [J. Phys. A: Math. Theor. 46 (2013) 494012 (34pp); Nuclear Physics B410 [FS] (1993) 535—549, J. Phys. A: Math. Theor. 46 (2013)].
%It was also suggested that, in such situations, the operator product expansion (OPE) should include logarithmic terms~\cite{VA02}. 
%Moreover, it was conjectured in [V. Guraire, Logarithmic operators in conformal field theory, Nuclear Physics B 410 (1993) 535-549] and [J. Phys. A: Math. Gen. 35 (2002) L377–L384] that new, so-called logarithmic, operators are responsible for the logarithmic terms appearing in the correlation functions and the OPE. 
%Since the work of Gurarie \cite{GURARIE1993535},
CFTs in which the correlators of basic fields can have logarithmic divergences at short distance are called logarithmic and have attracted considerable attention due to their role in the study of, e.g., the Wess-Zumino-Witten (WZW) model, the quantum Hall effect, disordered critical systems, self-avoiding polymers, percolation and the FK model.
However, despite significant recent progress~\cite{SKZ07,DV10,CR13,PSVD13,Dot16,Dot20,HGJS20,NR21,HS2022}, the field of logarithmic CFTs is considerably less developed than that of ordinary CFTs.
In particular, not many explicit examples of correlation functions exhibiting a logarithmic singularity are known.
%there is no general method to compute logarithmic correlation functions or to generate OPEs for logarithmic CFTs.

%Critical percolation was recognized early on as an important example of a logarithmic CFT, but the evidence for its logarithmic nature has been so far somewhat indirect.
For critical percolation in finite domains or the upper half-plane (i.e., in the presence of a boundary), logarithmic terms were identified in the expected number of clusters crossing a rectangle~\cite{Mai03} and in some crossing probabilities that generalize Cardy's crossing formula~\cite{SKZ07,Sim13}, and were conjectured~\cite{GV18} to appear in the boundary correlation function of four density fields, which we consider in Section \ref{sec:boundary}.
Our results rigorously confirm the logarithmic divergence found in~\cite{GV18}.

For critical percolation on the full plane, an example of a logarithmic correlation function was obtained in \cite{VJS12}, but in general, the study of the percolation CFT on the full plane (in the bulk) is considered much harder due to fact that the correlation functions of interest typically do not satisfy the BPZ equation and are therefore less constrained.
Our results of Section \ref{sec:bulk} are consistent with the conclusions of \cite{VJS12}, reached taking the limit $Q \to 1$ of Potts model correlation functions, and therefore provide an indirect test of the results and assumptions of \cite{VJS12}.
%and so our results support their conclusions as well as their assumptions (in particular, the fact that the $Q \to 1$ limit correctly reproduces the behavior of percolation correlations even though some of the Potts fields involved are ill-defined for $Q=1$).

%than in the presence of a boundary due to the constraints of crossing symmetry and the absence of null states (which implies that the correlation functions of interest do not satisfy the BPZ equation).
%Nevertheless, this formal limit is used in \cite{VJS12} to obtain an example of a logarithmic correlation function with a logarithmic divergence.
%This approach is used, for instance, in~\cite{VJS12} to obtain a limiting correlation function with a logarithmic divergence.
%\footnote{This example involves a non-primary field, since conformal invariance implies that the two-point function of primary fields should follow a pure power law.}
%As mentioned above, in \cite{VJS12}, a logarithmic term was found in the full-plane two-point function of an operator which is not a pure scaling operator,\footnote{Conformal covariance implies that the two-point function of a pure scaling operator is a pure power.} 

In both~\cite{GV18} and~\cite{VJS12}, the logarithmic terms are obtained by taking the limit $Q \to 1$ of correlation functions calculated for the $Q$-state Potts model with $Q \neq 1$.
A limitation of this approach is that the fields involved in the calculations are ill-defined for $Q=1$, while the logarithmic terms appear only in the singular limit $Q \to 1$.
This reveals the subtlety of the $Q \to 1$ limit and illustrates the problem of providing a direct calculation of a logarithmic correlation function in terms of well-defined percolation observables.

\section{Main results} \label{sec:main-results}

\subsection{The four-point function and OPE of the density field in the bulk} \label{sec:bulk}

The CFT description of two-dimensional critical percolation postulates the existence of a field $\psi(z)$ that measures the cluster density at a point $z$ and has scaling dimension\footnote{This suggests the identification $\psi=\phi_{0,1/2}$, where $\phi_{0,1/2}$ is the $Q \to 1$
	%$\psi=\phi_{3/2,3/2}$, where $\phi_{3/2,3/2}$ is the $Q \to 1$ $(m \to 2)$ 
	limit of the spin (magnetization) field %$\phi_{(m+1)/2,(m+1)/2}$ 
	of the Q-state Potts model \cite{DF84,Car84}, but $\psi(z)$ can be defined with no reference to the Potts model (see \cite{Camia24}).}
$2h_{\psi}=5/48$.
This field should satisfy
\begin{equation}
	\langle \psi(z_1)\psi(z_2) \rangle = P(z_1 \leftrightarrow z_2),
\end{equation}
where $P(z_1 \leftrightarrow z_2)$ is the ``probability'' that $z_1$ and $z_2$ belong to the same critical percolation cluster and $P$ should be interpreted as the continuum limit of the renormalized lattice probability, with a renormalization factor that scales like the inverse of the lattice spacing to the power of $8h_{\psi}$ (which means that renormalized ``probabilities'' can be larger than $1$).
A field with the properties described above is constructed in \cite{Camia24} and, following that construction, one has 
%The four-point function $\langle \psi(z_1)\psi(z_2)\psi(z_3)\psi(z_4) \rangle$ can be written as a sum of four terms (see \cite{Camia24}),
\begin{align} \label{eq:4-point-psi}
	\begin{split}
		& \langle \psi(z_1)\psi(z_2)\psi(z_3)\psi(z_4) \rangle = P(z_1 \leftrightarrow z_2 \leftrightarrow z_3 \leftrightarrow z_4) \\
		& \qquad + P(z_1 \leftrightarrow z_2 \not\leftrightarrow z_3 \leftrightarrow z_4) + P(z_1 \leftrightarrow z_3 \not\leftrightarrow z_2 \leftrightarrow z_4) + P(z_1 \leftrightarrow z_4 \not\leftrightarrow z_2 \leftrightarrow z_3),
	\end{split}
\end{align}
where the first term corresponds to the ``probability'' that all points belong to the same cluster and the remaining terms correspond to the ``probabilities'' that two points belong to one cluster and the remaining two to a different cluster. %\footnote{The function $P$ should be interpreted as the continuum limit of the renormalized lattice probability (the renormalization factor scales like the inverse of the lattice spacing to the power of $8h_{\psi}$) and can therefore be larger than $1$.}.
Asymptotically as $z_1,z_2 \to z$, we find that
\begin{align} \label{eq:4-point-psi-OPE}
	\begin{split}
		\langle \psi(z_1)\psi(z_2)\psi(z_3)\psi(z_4) \rangle \sim &\vert z_1-z_2 \vert^{-5/24} \Big[ \vert z_3-z_4 \vert^{-5/24}\\
& \qquad+ \vert z_1-z_2 \vert^{5/4} F(z,z_3,z_4) \big( C_0 - C_L \log\vert x \vert \big)\Big]
		%            & \qquad \qquad - C_L \log\Big(\frac{\vert z_3-z_4 \vert}{\vert z_1-z_4 \vert \vert z_3-z_2 \vert}\Big) \Big)\Big]
	\end{split}
\end{align}
for some finite constant $C_0$ and some $C_L \in (0,\infty)$, where
\begin{equation} \label{eq:def-F}
	F(z,z_3,z_4) = \vert z-z_3 \vert^{-5/4} \vert z-z_4 \vert^{-5/4} \vert z_3-z_4 \vert^{25/24},
\end{equation}
the cross-ratio
\begin{equation}
	x=\frac{(z_1-z_2)(z_3-z_4)}{(z_1-z_4)(z_3-z_2)}
\end{equation}
is invariant under M\"obius transformations, and $\log\vert x \vert \sim \log\vert z_1-z_2 \vert$.

%We obtain this result by a direct calculation, presented in Section \ref{sec:bulk_analysis}, which involves only percolation quantities.
%The novelty of our method is that it does not require to take the subtle $Q \to 1$ limit of Potts model quantities, not does it require any a priori assumptions on the operator content of the theory, providing the first direct calculation of a percolation four-point function in the bulk with a logarithmic singularity.
%In addition, our method can be turned into a mathematically rigorous derivation of \eqref{eq:4-point-psi-OPE}.

The four-point function \eqref{eq:4-point-psi-OPE} is consistent with the following logarithmic OPE:
\begin{align} \label{eq:logOPE}
	\begin{split}
		& \psi(z_1)\psi(z_2)  = \vert z_1-z_2 \vert^{-5/24} \Big[ 1 + \vert z_1-z_2 \vert^{5/4} \Big( \hat\phi(z) + \log\vert z_1-z_2 \vert \phi(z) \Big) + \ldots \Big],
	\end{split}
\end{align}
where $\hat\phi$ and $\phi$ are two distinct fields with scaling dimension $5/4$ such that
\begin{align} \label{eq:3-point-functions}
	\begin{split}
		& \langle \phi(z)\psi(z_3)\psi(z_4) \rangle = - C_L F(z,z_3,z_4) \\
		& \langle \hat\phi(z)\psi(z_3)\psi(z_4) \rangle = \Big( C_0 - C_L \log\Big(\frac{\vert z_3-z_4 \vert}{\vert z-z_4 \vert\vert z_3-z \vert}\Big)\Big) F(z,z_3,z_4) \\
		& \quad = \left( -\frac{C_0}{C_L} + \log\Big(\frac{\vert z_3-z_4 \vert}{\vert z-z_4 \vert\vert z_3-z \vert}\Big) \right) \langle \phi(z)\psi(z_3)\psi(z_4) \rangle.
	\end{split}
\end{align}

The field $\phi$ can be formally defined as
\begin{align} \label{eq:def-phi}
	\begin{split}
		& \phi(z)  = \lim_{\varepsilon \to 0} \frac{\varepsilon^{5/24} \psi(z)\psi(z+\varepsilon u) - 1}{\varepsilon^{5/4} \vert\log\varepsilon\vert},
	\end{split}
\end{align}
where $u$ is an arbitrary vector with $\vert u \vert = 1$.
Moreover, we have that
\begin{align} \label{eq:2-point-functions}
	\begin{split}
		& \langle\phi(z_1)\phi(z_2)\rangle=0 \\
		& \langle\phi(z_1)\hat\phi(z_2)\rangle = - C_L \vert z_1-z_2 \vert^{-5/2} \\
		& \langle \hat\phi(z_1)\hat\phi(z_2) \rangle = \frac{C_0 + 2 C_L \log\vert z_1-z_2 \vert}{\vert z_1-z_2 \vert^{5/2}} \, .
	\end{split}
\end{align}

Equations \eqref{eq:2-point-functions} identify $\phi$ and $\hat\phi$ as a logarithmic pair \cite{Gur93,CR13}.
It is interesting to note that the first of the equations \eqref{eq:2-point-functions} is a direct consequence of the logarithm in the definition of $\phi$ \eqref{eq:def-phi}.
%have the form predicted by Gurarie \cite{Gur93,CR13} for a consistent CFT with central charge $c=0$ and confirm the presence of a logarithmic pair of operators in the CFT of percolation in the bulk.
The field $\phi$ can be naturally identified with (a multiple of) the so-called energy field (a conformal primary) discussed in Section~4.2 of~\cite{Car13} and $\hat\phi$ with its logarithmic partner.

%It seems natural to identify $\phi$ and $\hat\phi$ with the fields discussed in Section~4.2 of~\cite{Car13}, namely, $\phi$ with the energy operator and $\hat\phi$ with %its logarithmic partner, 
%the so-called two-cluster operator.

\subsection{The four-point function and OPE of the density field on the boundary} \label{sec:boundary}

Here we focus on critical percolation in the upper half-plane, with the real line as the boundary, and consider the four-point function $\langle\phi_{1,3}(x_1)\phi_{1,3}(x_2)\phi_{1,3}(x_3)\phi_{1,3}(x_4)\rangle$, where $x_1<x_2<x_3<x_4$ are points on the real line, and the insertion of $\phi_{1,3}$ corresponds to anchoring a macroscopic cluster to a point on the boundary, as discussed in~\cite{JohnCardy_1998,PhysRevLett.97.115702}.
Consequently, the four-point function corresponds to a sum of renormalized probabilities, as follows:
\begin{align} \label{eq:4-point-phi}
	\begin{split}
		& \langle \phi_{1,3}(x_1)\phi_{1,3}(x_2)\phi_{1,3}(x_3)\phi_{1,3}(x_4) \rangle = P(x_1 \leftrightarrow x_2 \leftrightarrow x_3 \leftrightarrow x_4) \\
		& \qquad + P(x_1 \leftrightarrow x_2 \not\leftrightarrow x_3 \leftrightarrow x_4) + P(x_1 \leftrightarrow x_4 \not\leftrightarrow x_2 \leftrightarrow x_3),
	\end{split}
\end{align}
where the first term corresponds to the ``probability'' that all points belong to the same cluster and the remaining terms correspond to the ``probabilities'' that two points belong to one cluster and the remaining two to a different cluster.
(Note that, since $x_1<x_2<x_3<x_4$, the event $x_1 \leftrightarrow x_3 \not\leftrightarrow x_2 \leftrightarrow x_4$ is not possible on the upper half-plane for topological reasons.)
From a CFT perspective, the field $\phi_{1,3}$ has conformal dimension $h_{1,3}=1/3$, which coincides with the value of the percolation boundary one-arm exponent \cite{ADA99,SW01}, and appears in the OPE
\begin{equation} \label{eq:OPE_phi12}
	\phi_{1,2} \times \phi_{1,2} = 1 + \phi_{1,3} + \ldots,
\end{equation}
where the field $\phi_{1,2}$, with conformal dimension $h_{1,2}=0$, corresponds to Cardy's boundary condition changing operator, which implements a change of boundary condition at the insertion point, and can be identified with the $O(N=1)$ one-leg operator inserting a single chordal SLE$_6$ \cite{ADA99}.

As $x_1,x_2 \to x < x_3$, a calculation similar to that leading to \eqref{eq:4-point-psi-OPE} gives
\begin{align} \label{eq:4-point-psi-OPE-half-plane}
	\begin{split}
		& \langle\phi_{1,3}(x_1)\phi_{1,3}(x_2)\phi_{1,3}(x_3)\phi_{1,3}(x_4)\rangle \sim (x_2-x_1)^{-2/3} \Bigg[ (x_4-x_3)^{-2/3} \\
		& \qquad + (x_2-x_1)^{2} F_H(x,x_3,x_4) \Big( C_{H,0} - C_{H,L} \log\Big\vert \frac{(x_1-x_2)(x_3-x_4)}{(x_1-x_4)(x_3-x_2)} \Big\vert \Big)\Bigg]
	\end{split}
\end{align}
for some finite constant $C_{H,0}$ and some $C_{H,L} \in (0,\infty)$, where
\begin{equation} \label{eq:def-F_H}
    F_H(x,x_3,x_4) = (x_3-x)^{-2} (x_4-x)^{-2} (x_4-x_3)^{4/3}.
\end{equation}

The four-point function \eqref{eq:4-point-psi-OPE-half-plane} is consistent with the following logarithmic OPE:
\begin{align} \label{eq:half-plane-logOPE}
	\begin{split}
		& \phi_{1,3}(x_1)\phi_{1,3}(x_2) = (x_2-x_1)^{-2/3} \Big[ 1 + (x_2-x_1)^{2} \Big( t(x) + \log(x_2-x_1) T(x) \Big) + \ldots \Big],
	\end{split}
\end{align}
where $t$ and $T$ are two distinct fields of scaling dimension $2$ such that
\begin{align} \label{eq:half-plane-3-point-functions}
	\begin{split}
		& \langle T(x)\phi_{1,3}(x_3)\phi_{1,3}(x_4) \rangle = -C_{H,L} F_H(x,x_3,x_4) \\
		& \langle t(x)\phi_{1,3}(x_3)\phi_{1,3}(x_4) \rangle = \Big( C_{H,0} - C_{H,L} \log\Big(\frac{(x_4-x_3)}{(x_4-x)(x_3-x)}\Big) \Big) F_H(x,x_3,x_4) \\
		& \quad = \left( -\frac{C_{H,0}}{C_{H,L}} + \log\Big(\frac{(x_4-x_3)}{(x_4-x)(x_3-x)}\Big) \right) \langle T(x)\phi_{1,3}(x_3)\phi_{1,3}(x_4) \rangle .
	\end{split}
\end{align}

The field $T$ can be formally defined as
\begin{align} \label{eq:def-T}
	\begin{split}
		& T(x)  = \lim_{\varepsilon \to 0} \frac{\varepsilon^{2/3} \phi_{1,3}(x)\phi_{1,3}(x+\varepsilon) - 1}{\varepsilon^{2} \vert\log\varepsilon\vert}.
	\end{split}
\end{align}
Moreover, we have that
\begin{align} \label{eq:2-point-functions-boundary}
	\begin{split}
		& \langle T(x_1) T(x_2)\rangle=0 \\
		& \langle T(x_1) t(x_2) \rangle = -C_{H,L} \vert x_1-x_2 \vert^{-4} \\
		& \langle t(x_1)t(x_2) \rangle = \frac{C_{H,0} + 2 C_{H,L} \log\vert x_1-x_2 \vert}{\vert x_1-x_2 \vert^{4}} \, .
	\end{split}
\end{align}

%Moreover, we have that
%\begin{equation} \label{eq:<TT>}
%	\langle T(x_1) T(x_2)\rangle=0
%\end{equation}
%and
%\begin{equation} \label{eq:<tT>}
%	\langle t(x_1) T(x_2)\rangle = C_{\phi_{1,3}\phi_{1,3}T} \vert x_1-x_2 \vert^{-4},
%\end{equation}
%where $C_{\phi_{1,3}\phi_{1,3}T}$ is the structure constant appearing in the 3-point function
%\begin{equation}
%	\langle \phi_{1,3}(x_1) \phi_{1,3}(x_2) T(x_3) \rangle = C_{\phi_{1,3}\phi_{1,3}T} \vert x_1-x_2 \vert^{4/3} \vert x_1-x_3 \vert^{-2} \vert x_2-x_3 \vert^{-2}.
%\end{equation}

The field $T$ can be identified with the stress-energy tensor of the theory and $t$ with its logarithmic partner.
As predicted by Gurarie \cite{Gur93}, the presence of a logarithmic partner to the stress-energy tensor is linked to the appearance of the logarithmic terms in the 4-point-function \eqref{eq:4-point-psi-OPE-half-plane} and the OPE \eqref{eq:half-plane-logOPE}.

\subsection{A mixed four-point function of boundary fields} \label{sec:mixed}

For critical percolation on the upper half-plane, the boundary three-leg operator $\phi_{1,4}(x)$ corresponds to the insertion of three interfaces near $x$ (leading to three chordal SLE$_6$ curves in the continuum limit) or, equivalently, to the creation of two macroscopic clusters (see, for example, \cite{ADA99}).
The field $\phi_{1,4}$ has conformal dimension $h_{1,4}=1$, which coincides with the value of the percolation boundary two-arm exponent \cite{ADA99,SW01}.

Given four points on the real line, $-\infty<x_1<x_2<x_3<x_4<\infty$, the four-point function $\langle \phi_{1,4}(x_1) \phi_{1,4}(x_2) \phi_{1,2}(x_3) \phi_{1,2}(x_4) \rangle$ corresponds to the insertion of three interfaces (chordal SLE$_6$ curves, in the continuum limit) near (at, in the continuum limit) $x_1$ and $x_2$ and one interface (chordal SLE$_6$) near (at) $x_3$ and $x_4$, as represented in Fig.~\ref{fig:SLE_insertion}.
The interfaces can be interpreted as defect lines in the $O(N=1)$ model, at least at the level of scaling dimensions (see, e.g., \cite{ADA99}).

A careful analysis of the configurations contributing to the four-point function leads to
\begin{align}
	\langle \phi_{1,4}(x_1) \phi_{1,4}(x_2) \phi_{1,2}(x_3) \phi_{1,2}(x_4) \rangle = (x_2-x_1)^{-2} \big( 1 + f_0(x) + C \log x \big),
	%\vert x_1-x_2 \vert^{-2} \big( 1 + \hat{C}_L \log x + \hat{C}_0 (x_2-x_1)^{1/3} + O\big((x_2-x_1)^{2}\big)\big),
\end{align}
%as $x_2-x_1 \to 0$,
for some constant $C \in (0,\infty)$, where
\begin{equation} \label{cross-ratio}
	x=\frac{(x_4-x_2)(x_3-x_1)}{(x_3-x_2)(x_4-x_1)}
\end{equation}
is the conformally-invariant cross-ratio and the function $f_0$ satisfies
\begin{itemize}
	\item $f_0(x) \sim (x_2-x_1)^{1/3}+O\big((x_2-x_1)^2\big)$ as $x_2-x_1 \to 0$,
	\item $f_0(x) \sim (x_4-x_3)^{1/3}+O(x_4-x_3)$ as $x_4-x_3 \to 0$,
	\item $f_0(x) \sim (x_3-x_2)^{-2/3}$ as $x_3-x_2 \to 0$.
\end{itemize}

\section{Derivations of the main results}

\subsection{Analysis of the four-point function of the density field in the bulk} \label{sec:bulk_analysis}

In this subsection, we analyze the four-point function $\langle \psi(z_1)\psi(z_2)\psi(z_3)\psi(z_4) \rangle = P(z_1 \leftrightarrow z_2 \leftrightarrow z_3 \leftrightarrow z_4)$.
From~\eqref{eq:4-point-psi} one can see that, when $z_1,z_2 \to z$, the leading behavior comes from the first two terms in the right-hand side of \eqref{eq:4-point-psi} and is given by $\vert z_1-z_2 \vert^{-5/24} \vert z_3-z_4 \vert^{-5/24}$.
The last two terms require that the clusters of $z_1$ and $z_2$ be different.
Therefore, as $z_1,z_2 \to z$, they correspond to the insertion at $z$ of an operator $\phi$ producing two distinct clusters, which is related to the four-leg operator with scaling dimension $5/4$, where the four legs are produced by the boundaries of the two clusters.
Hence, the leading contribution of the last two terms is $\vert z_1-z_2 \vert^{-5/24} \vert z_1-z_2 \vert^{5/4} F(z,z_3,z_4)$,
%\begin{align}
%    \vert z_1-z_2 \vert^{-5/24} \vert z_1-z_2 \vert^{5/4} F(z,z_3,z_4),
%\end{align}
where $F(z,z_3,z_4) = \vert z-z_3 \vert^{-5/4} \vert z-z_4 \vert^{-5/4} \vert z_3-z_4 \vert^{25/24}$.
As observed in~\cite{Dot16} and \cite{Camia24}, this suggests the OPE
\begin{align} \label{eq:OPE}
	\begin{split}
		\psi(z_1)\psi(z_2)  = \vert z_1-z_2 \vert^{-5/24} \big( 1 + C_{\psi\psi\phi} \vert z_1-z_2 \vert^{5/4} \phi(z) + \ldots \big),
	\end{split}
\end{align}
where $C_{\psi\psi\phi}$ is the structure constant appearing in the three-point function $\langle \psi(z_1)\psi(z_2)\phi(z_3)\rangle$ and the ellipsis denotes the contribution from other operators.
%$\phi_{2,0}$ is the four-leg operator, and an asymptotic behavior for the four-point function involving, at least at first and second order, only power laws.

%In the next section,
We will now show that a more careful analysis of the first two terms in the right-hand side of \eqref{eq:4-point-psi} reveals the presence of a term that behaves like $\vert z_1-z_2 \vert^{-5/24} \vert z_1-z_2 \vert^{5/4} F(z,z_3,z_4) \big\vert\log \vert z_1-z_2 \vert\big\vert$.

\begin{figure}
	\includegraphics[width= 0.4\textwidth]{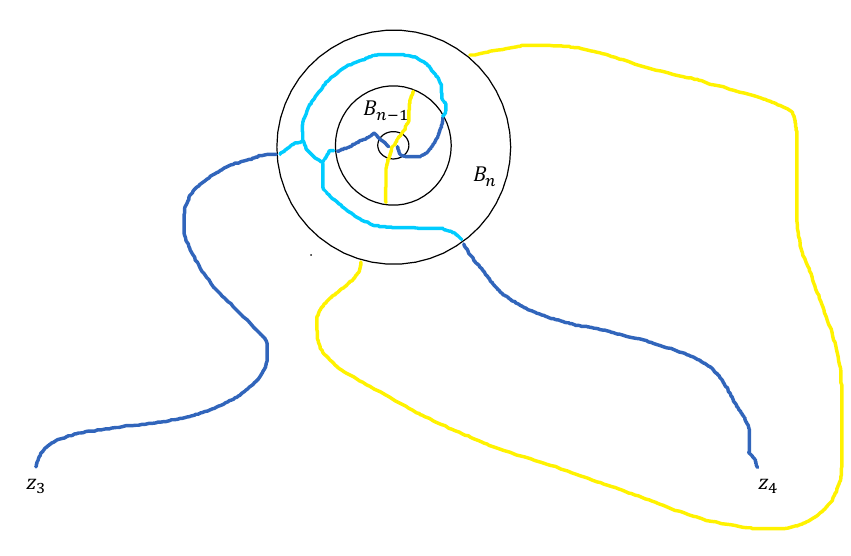}
	\caption{The event $\{ z_1 \xleftrightarrow[B_{n-1}]{B_n} z_2, z_3 \xleftrightarrow[B_n^c]{} z_4 \}$. Yellow lines (lightest shade) denote closed paths, (light) blue lines (darker shades) denote open paths. $z_1$ and $z_2$ are contained in the inner disk and not marked. They are not connected by an open path within the next disk, $B_{n-1}$, but are connected within the largest disk, $B_n$, with radius twice that of $B_{n-1}$. $z_3$ and $z_4$ are connected by an open path, but not outside $B_n$. The open paths connecting $z_1,z_2$ and $z_3,z_4$ can overlap within $B_n$. The number, $N$, of disks one can insert between $z_1,z_2$ and $z_3,z_4$ is of order $\log{(1/\vert z_1-z_2 \vert)}$.}
	\label{Figure}
\end{figure}

%\section{Logarithmic singularity}
In describing percolation events, we will use the standard terminology of open and closed clusters and paths.
Given two subsets of the plane, $A$ and $B$, we consider the following events:
\begin{itemize}
	\item $z_1 \xlongleftrightarrow[]{A} z_2$: there is an open path between $z_1$ and $z_2$ contained in $A$,
	\item $z_1 \xlongleftrightarrow[B]{} z_2$: $z_1,z_2$ belong to the same open cluster but there is no open path fully contained in $B$,
	\item $z_1 \xlongleftrightarrow[B]{A} z_2$: there is an open path between $z_1$ and $z_2$ contained in $A$ but no open path fully contained in $B$.
\end{itemize}
Now consider disks $B_n = \{ z : \vert z - \frac{z_1+z_2}{2} \vert \leq 2^n \vert z_1-z_2 \vert\}$ for $n=1,\ldots,N$, where $N$ is chosen so that $2^N \sim 1/\vert z_1-z_2 \vert$, that is, $N \sim -\log \vert z_1-z_2 \vert$.
We assume that $z_1$ and $z_2$ are close to each other and that $z_3$ and $z_4$ are outside $B_N$ (see Fig.~\ref{Figure}).
Letting $B_n^c=\mathbb{C} \setminus B_n$ denote the complement of $B_n$, the sum of the first two terms of \eqref{eq:4-point-psi} can be written as
\begin{align} \label{eq:summation}
	\begin{split}
		%& P(z_1 \leftrightarrow z_2 \leftrightarrow z_3 \leftrightarrow z_4) + P(z_1 \leftrightarrow z_2 \not\leftrightarrow z_3 \leftrightarrow z_4) = 
		& P(z_1 \leftrightarrow z_2, z_3 \leftrightarrow z_4) \\
		&= P(z_1 \xleftrightarrow[]{B_1} z_2, z_3 \leftrightarrow z_4) + \sum_{n=2}^{N} P(z_1 \xleftrightarrow[B_{n-1}]{B_n} z_2, z_3 \leftrightarrow z_4) + P(z_1 \xleftrightarrow[B_N]{} z_2, z_3 \leftrightarrow z_4) \\
		& = P\Big(z_1 \xleftrightarrow[]{B_1} z_2, z_3 \xleftrightarrow[]{B_1^c} z_4 \Big) + P(z_1 \xleftrightarrow[]{B_1} z_2, z_3 \xleftrightarrow[B_1^c]{} z_4) \\
		& \quad + \sum_{n=2}^{N} \Big[ P(z_1 \xleftrightarrow[B_{n-1}]{B_n} z_2, z_3 \xleftrightarrow[]{B_n^c} z_4) + P(z_1 \xleftrightarrow[B_{n-1}]{B_n} z_2, z_3 \xleftrightarrow[B_n^c]{} z_4)\Big] + P(z_1 \xleftrightarrow[B_N]{} z_2, z_3 \leftrightarrow z_4) \\
		& = P(z_1 \xleftrightarrow[]{B_1} z_2) P(z_3 \xleftrightarrow[]{B_1^c} z_4) + P(z_1 \xleftrightarrow[]{B_1} z_2, z_3 \xleftrightarrow[B_1^c]{} z_4) \\
		& \quad + \sum_{n=2}^{N} \Big[P(z_1 \xleftrightarrow[B_{n-1}]{B_n} z_2) P(z_3 \xleftrightarrow[]{B_n^c} z_4) + P(z_1 \xleftrightarrow[B_{n-1}]{B_n} z_2, z_3 \xleftrightarrow[B_n^c]{} z_4)\Big] + P(z_1 \xleftrightarrow[B_N]{} z_2, z_3 \leftrightarrow z_4),
	\end{split}
\end{align}
where the last equality follows from the independence of the percolation events considered.
Similarly, we have
{\small \begin{align} \label{eq:product}
		\begin{split}
			& P(z_1 \leftrightarrow z_2) P(z_3 \leftrightarrow z_4)
			%= \Big[ P(z_1 \xleftrightarrow[]{B_1} z_2) \\
			%			& \quad + \sum_{n=2}^{N} P(z_1 \xleftrightarrow[B_{n-1}]{B_n} z_2) + P(z_1 \xleftrightarrow[B_N]{} z_2) \Big] P(z_3 \leftrightarrow z_4) \\
			= P(z_1 \xleftrightarrow[]{B_1} z_2) P(z_3 \xleftrightarrow[]{B_1^c} z_4) + P(z_1 \xleftrightarrow[]{B_1} z_2) P(z_3 \xleftrightarrow[B_1^c]{} z_4) \\
			& \quad + \sum_{n=2}^{N} \Big[ P(z_1 \xleftrightarrow[B_{n-1}]{B_n} z_2) P(z_3 \xleftrightarrow[]{B_n^c} z_4) + P(z_1 \xleftrightarrow[B_{n-1}]{B_n} z_2) P(z_3 \xleftrightarrow[B_n^c]{} z_4) \Big] + P(z_1 \xleftrightarrow[B_N]{} z_2) P(z_3 \leftrightarrow z_4).
		\end{split}
\end{align}}

Comparing \eqref{eq:summation} and \eqref{eq:product}, we obtain
\begin{align} \label{eq:compare}
	\small
	\begin{split}
		& P(z_1 \leftrightarrow z_2, z_3 \leftrightarrow z_4) \\
		& = P(z_1 \leftrightarrow z_2) P(z_3 \leftrightarrow z_4) + \Big[ P(z_3 \xleftrightarrow[B_1^c]{} z_4 \vert z_1 \xleftrightarrow[]{B_1} z_2) - P(z_3 \xleftrightarrow[B_1^c]{} z_4) \Big]  P(z_1 \xleftrightarrow[]{B_1} z_2) \\
		& + \sum_{n=2}^{N} \Big[ P(z_1 \xleftrightarrow[B_{n-1}]{} z_2, z_3 \xleftrightarrow[B_n^c]{} z_4 \vert z_1 \xleftrightarrow[]{B_n} z_2) - P(z_1 \xleftrightarrow[B_{n-1}]{} z_2 \vert z_1 \xleftrightarrow[]{B_n} z_2) P(z_3 \xleftrightarrow[B_n^c]{} z_4) \Big] P(z_1 \xleftrightarrow[]{B_n} z_2) \\
		& + \Big[ P(z_1 \xleftrightarrow[B_N]{} z_2, z_3 \leftrightarrow z_4 \vert z_1 \leftrightarrow z_2) - P(z_1 \xleftrightarrow[B_N]{} z_2 \vert z_1 \leftrightarrow z_2) P(z_3 \leftrightarrow z_4) \Big] P(z_1 \leftrightarrow z_2)
	\end{split}
\end{align}
and we see that the leading behavior comes from the term $P(z_1 \leftrightarrow z_2) P(z_3 \leftrightarrow z_4) \sim \vert z_1-z_2 \vert^{-5/24} \vert z_3-z_4 \vert^{-5/24}$.

To understand the behavior of the next term, observe that the event $z_3 \xleftrightarrow[B_1^c]{} z_4$ implies that, in $B^c_1$, $z_3$ and $z_4$ are connected to $\partial B_1$ by two disjoint open clusters which are separated by a closed cluster. We denote this event by $\mathcal{F}(z_3,z_4;B_1^c)$.
Since $\mathcal{F}(z_3,z_4;B_1^c)$ means that the annulus $B_N \setminus B_1$ is crossed by two open paths (connecting $z_3$ and $z_4$ to $\partial B_1$) and two closed paths (which must be present to separate the open clusters supporting the two open paths),
%prevent the existence of an open path between $z_3$ and $z_4$ fully contained in $B_1^c$) %which implies that it is crossed by four interfaces (cluster boundaries).
the four-arm exponent \cite{ADA99,SW01}
implies that $P(\mathcal{F}(z_3,z_4;B_1^c)) \sim \vert z_1-z_2 \vert^{5/4}$.
Conditioning on $\mathcal{F}(z_3,z_4;B_1^c)$, we can write
\begin{align} \label{eq:FB1}
	\begin{split}
		%        & P(z_1 \xleftrightarrow[]{B_1} z_2, z_3 \xleftrightarrow[B_1^c]{} z_4) -  P(z_1 \xleftrightarrow[]{B_1} z_2) P(z_3 \xleftrightarrow[B_1^c]{} z_4) \\
		%        & \quad = \big[P(z_1 \xleftrightarrow[]{B_1} z_2, z_3 \xleftrightarrow[B_1^c]{} z_4 \vert \mathcal{F}(z_3,z_4;B_1^c) -  P(z_1 \xleftrightarrow[]{B_1} z_2) P(z_3 \xleftrightarrow[B_1^c]{} z_4 \vert \mathcal{F}(z_3,z_4;B_1^c))\big] P(\mathcal{F}(z_3,z_4;B_1^c)) \\
		& \big[ P(z_3 \xleftrightarrow[B_1^c]{} z_4 \vert z_1 \xleftrightarrow[]{B_1} z_2) - P(z_3 \xleftrightarrow[B_1^c]{} z_4) \big]  P(z_1 \xleftrightarrow[]{B_1} z_2) \\
		& = \big[P(z_3 \xleftrightarrow[B_1^c]{} z_4 \vert z_1 \xleftrightarrow[]{B_1} z_2, \mathcal{F}(z_3,z_4;B_1^c)) - P(z_3 \xleftrightarrow[B_1^c]{} z_4 \vert \mathcal{F}(z_3,z_4;B_1^c))\big] \\
		&\quad\times P(z_1 \xleftrightarrow[]{B_1} z_2) P(\mathcal{F}(z_3,z_4;B_1^c)),
	\end{split}
\end{align}
where $P(z_1 \xleftrightarrow[]{B_1} z_2) P(\mathcal{F}(z_3,z_4;B_1^c)) \sim \vert z_1-z_2 \vert^{-5/24} \vert z_1-z_2 \vert^{5/4}$.
In a configuration such that $\mathcal{F}(z_3,z_4;B_1^c)$ occurs, the event $z_3 \xlongleftrightarrow[B_1^c]{} z_4$ reduces to the event that the clusters of $z_3$ and $z_4$, which are disjoint when restricted to $B^c_1$, are connected inside $B_1$.
Conditioning on $z_1 \xlongleftrightarrow[]{B_1} z_2$ makes it easier for such an event to happen, so
%that
%\begin{equation} \label{eq:next-term}
%\begin{split}
%		&P(z_3 \xleftrightarrow[B_1^c]{} z_4 \vert z_1 \xleftrightarrow[]{B_1} z_2, \mathcal{F}(z_3,z_4;B_1^c)) \\
%	&\quad-P(z_3 \xleftrightarrow[B_1^c]{} z_4 \vert \mathcal{F}(z_3,z_4;B_1^c)) \geq 0.
%\end{split}
%\end{equation}
we can conclude that
\begin{equation} \label{eqn::innermost_term}
	\begin{split}
		%P(z_1 \xleftrightarrow[]{B_1} z_2, z_3 \xleftrightarrow[B_1^c]{} z_4) -  P(z_1 \xleftrightarrow[]{B_1} z_2) P(z_3 \xleftrightarrow[B_1^c]{} z_4)
		& \big[ P(z_3 \xleftrightarrow[B_1^c]{} z_4 \vert z_1 \xleftrightarrow[]{B_1} z_2) - P(z_3 \xleftrightarrow[B_1^c]{} z_4) \big]  P(z_1 \xleftrightarrow[]{B_1} z_2) \sim \vert z_1-z_2 \vert^{-5/24} \vert z_1-z_2 \vert^{5/4}.
	\end{split}
\end{equation}

For $n>1$ (see Fig.~\ref{Figure}), the event $z_3 \xleftrightarrow[B_n^c]{} z_4$ is analogous to $z_3 \xleftrightarrow[B_1^c]{} z_4$ and induces the event $\mathcal{F}(z_3,z_4;B_n^c)$ such that $P(\mathcal{F}(z_3,z_4;B_n^c)) \sim \big( 2^n \vert z_1-z_2 \vert \big)^{5/4}$.
Furthermore, the event $z_1 \xleftrightarrow[B_{n-1}]{} z_2$ implies that, inside $B_{n-1}$, $z_1$ and $z_2$ are connected to $\partial B_{n-1}$ by two disjoint open clusters, separated by a closed cluster. We denote this event by $\mathcal{F}(z_1,z_2;B_{n-1})$.
Since $\mathcal{F}(z_1,z_2;B_{n-1})$ implies that the annulus $B_{n-1} \setminus B_1$ is crossed by two disjoint open paths and two disjoint closed paths, the four-arm exponent gives $P(\mathcal{F}(z_1,z_2;B_{n-1})) \sim \big((1/2)^{n-2}\big)^{5/4}$.
Using the fact that $B_{n-1}$ and $B_n^c$ are disjoint, we see that 
\begin{equation} \label{eq:product-of-powers}
	\begin{split}
		& P(\mathcal{F}(z_1,z_2;B_{n-1}), \mathcal{F}(z_3,z_4;B_n^c)) = P(\mathcal{F}(z_1,z_2;B_{n-1})) P(\mathcal{F}(z_3,z_4;B_n^c)) \sim \vert z_1-z_2 \vert^{5/4},
	\end{split}
\end{equation}
independently of $n$.
With this, using arguments analogous to those above, we can write
\begin{align}
	\begin{split} \label{eq:difference}
		& \Big[ P(z_1 \xleftrightarrow[B_{n-1}]{} z_2, z_3 \xleftrightarrow[B_n^c]{} z_4 \vert z_1 \xleftrightarrow[]{B_n} z_2) - P(z_1 \xleftrightarrow[B_{n-1}]{} z_2 \vert z_1 \xleftrightarrow[]{B_n} z_2) P(z_3 \xleftrightarrow[B_n^c]{} z_4) \Big] P(z_1 \xleftrightarrow[]{B_n} z_2) \\
		& \quad = \Big[ P(z_1 \xleftrightarrow[B_{n-1}]{} z_2, z_3 \xleftrightarrow[B_n^c]{} z_4 \vert z_1 \xleftrightarrow[]{B_n} z_2, \mathcal{F}(z_1,z_2;B_{n-1}), \mathcal{F}(z_3,z_4;B_n^c)) \\
		& \qquad - P(z_1 \xleftrightarrow[B_{n-1}]{} z_2 \vert z_1 \xleftrightarrow[]{B_n} z_2, \mathcal{F}(z_1,z_2;B_{n-1})) P(z_3 \xleftrightarrow[B_n^c]{} z_4 \vert \mathcal{F}(z_3,z_4;B_n^c)) \Big] \\
		& \qquad \quad P(z_1 \xleftrightarrow[]{B_n} z_2) P(\mathcal{F}(z_1,z_2;B_{n-1})) P(\mathcal{F}(z_3,z_4;B_n^c)) \\
		& \quad \sim g_n(z_1,z_2,z_3,z_4) \vert z_1-z_2 \vert^{-5/24} \vert z_1-z_2 \vert^{5/4},
	\end{split}
\end{align}
%where we introduced the notation
%{\color{red}{
		%\begin{widetext}
		%		\begin{align}
			%		\begin{split} \label{eq:positive-difference}
				%			g_n(z_1,z_2,z_3,z_4) = P(z_3 \xleftrightarrow[B_n^c]{} z_4 \vert z_1 \xleftrightarrow[B_{n-1}]{B_n} z_2, \mathcal{F}(z_3,z_4;B_n^c)) - P(z_3 \xleftrightarrow[B_n^c]{} z_4 \vert \mathcal{F}(z_3,z_4;B_n^c)).
				%   P(z_1 \xleftrightarrow[B_{n-1}]{} z_2, z_3 \xleftrightarrow[B_n^c]{} z_4 \vert z_1 \xleftrightarrow[]{B_n} z_2, \mathcal{F}(z_1,z_2;B_{n-1}), \mathcal{F}(z_3,z_4;B_n^c)) \\
				%			& \qquad - P(z_1 \xleftrightarrow[B_{n-1}]{} z_2 \vert z_1 \xleftrightarrow[]{B_n} z_2, \mathcal{F}(z_1,z_2;B_{n-1})) P(z_3 \xleftrightarrow[B_n^c]{} z_4 \vert \mathcal{F}(z_3,z_4;B_n^c)).
				%		\end{split}
			%	\end{align}
		%\end{widetext}
		%}
	where we introduced the notation
	\begin{align}
		\begin{split} \label{eq:positive-difference}
			& g_n(z_1,z_2,z_3,z_4) = P(z_3 \xleftrightarrow[B_n^c]{} z_4 \vert z_1 \xleftrightarrow[B_{n-1}]{B_n} z_2, \mathcal{F}(z_3,z_4;B_n^c)) - P(z_3 \xleftrightarrow[B_n^c]{} z_4 \vert \mathcal{F}(z_3,z_4;B_n^c)).
			%   P(z_1 \xleftrightarrow[B_{n-1}]{} z_2, z_3 \xleftrightarrow[B_n^c]{} z_4 \vert z_1 \xleftrightarrow[]{B_n} z_2, \mathcal{F}(z_1,z_2;B_{n-1}), \mathcal{F}(z_3,z_4;B_n^c)) \\
			%			& \qquad - P(z_1 \xleftrightarrow[B_{n-1}]{} z_2 \vert z_1 \xleftrightarrow[]{B_n} z_2, \mathcal{F}(z_1,z_2;B_{n-1})) P(z_3 \xleftrightarrow[B_n^c]{} z_4 \vert \mathcal{F}(z_3,z_4;B_n^c)).
		\end{split}
	\end{align}
	
	From scale invariance and the fact that different connectivity events inside $B_n \setminus B_{n-1}$ can use the same open paths and therefore ``help each other'' (see Fig.~\ref{Figure}), one can deduce \cite{CF24} that $g_n(z_1,z_2,z_3,z_4)$ is positive and bounded away from zero for all $n$ of order $-\log\vert z_1-z_2 \vert$, for which the disks $B_n$ and $B_{n-1}$ are macroscopic.
	Therefore,
	\begin{align}
		\begin{split} \label{eq:log}
			& \sum_{n=2}^{N} \big[ P(z_1 \xleftrightarrow[B_{n-1}]{} z_2, z_3 \xleftrightarrow[B_n^c]{} z_4 \vert z_1 \xleftrightarrow[]{B_n} z_2) - P(z_1 \xleftrightarrow[B_{n-1}]{} z_2 \vert z_1 \xleftrightarrow[]{B_n} z_2) P(z_3 \xleftrightarrow[B_n^c]{} z_4) \big] P(z_1 \xleftrightarrow[]{B_n} z_2) \\
			& \quad \sim - g(z_1,z_2,z_3,z_4) \vert z_1-z_2 \vert^{-5/24} \vert z_1-z_2 \vert^{5/4} \log \vert z_1-z_2 \vert,
		\end{split}
	\end{align}
	for some $g(z_1,z_2,z_3,z_4)>0$, which shows that $P(z_1 \longleftrightarrow z_2, z_3 \longleftrightarrow z_4)$ contains a term with a logarithmic divergence as $\vert z_1-z_2 \vert \to 0$.
	Similar considerations imply that the remaining term in the expression \eqref{eq:compare} of $P(z_1 \longleftrightarrow z_2, z_3 \longleftrightarrow z_4)$ behaves like
	\begin{align}
		\small
		\begin{split} \label{eq:last-term}
			& \big[ P(z_1 \xleftrightarrow[B_N]{} z_2, z_3 \leftrightarrow z_4 \vert z_1 \leftrightarrow z_2) - P(z_1 \xleftrightarrow[B_N]{} z_2 \vert z_1 \leftrightarrow z_2) P(z_3 \leftrightarrow z_4) \big] P(z_1 \leftrightarrow z_2) \\
			&  = \big[ P(z_1 \xleftrightarrow[B_N]{} z_2, z_3 \leftrightarrow z_4 \vert z_1 \leftrightarrow z_2, \mathcal{F}(z_1,z_2;B_N)) - P(z_1 \xleftrightarrow[B_N]{} z_2 \vert z_1 \leftrightarrow z_2, \mathcal{F}(z_1,z_2;B_N)) P(z_3 \leftrightarrow z_4) \big] \\
			& \quad \quad P(z_1 \leftrightarrow z_2) P(\mathcal{F}(z_1,z_2;B_N)) \sim \vert z_1-z_2 \vert^{-5/24} \vert z_1-z_2 \vert^{5/4}.
		\end{split}
	\end{align}
	
	We now write \eqref{eq:4-point-psi} as
	\begin{align}
		\begin{split}
			& \langle \psi(z_1)\psi(z_2)\psi(z_3)\psi(z_4) \rangle = P(z_1 \leftrightarrow z_2) P(z_3 \leftrightarrow z_4) + G(z_1,z_2,z_3,z_4) \\
			& \quad + P(z_1 \leftrightarrow z_3 \not\leftrightarrow z_2 \leftrightarrow z_4) + P(z_1 \leftrightarrow z_4 \not\leftrightarrow z_2 \leftrightarrow z_3),
		\end{split}
	\end{align}
	where
	\begin{align}
		\begin{split}
			& G(z_1,z_2,z_3,z_4) = P(z_1 \longleftrightarrow z_2, z_3 \longleftrightarrow z_4) - P(z_1 \longleftrightarrow z_2) P(z_3 \longleftrightarrow z_4)
		\end{split}
	\end{align}
	involves the insertion of four density operators and therefore,\footnote{One can see this, for example, using arguments from \cite{Camia24}.} if $f$ is a M\"obius transformation,
	{\begin{align}
			\begin{split}
				&G(f(z_1),f(z_2),f(z_3),f(z_4)) = \prod_{j=1}^4 \vert f'(z_j) \vert^{-5/48} G(z_1,z_2,z_3,z_4).
			\end{split}
	\end{align}}
	Combined with \eqref{eq:compare}, \eqref{eqn::innermost_term}, \eqref{eq:log} and \eqref{eq:last-term}, this implies that, when $z_1,z_2 \to z$,
	\begin{equation} \label{eq:G}
		\begin{split}
			&G(z_1,z_2,z_3,z_4) \sim - F(z,z_3,z_4) \vert z_1-z_2 \vert^{-5/24} \vert z_1-z_2 \vert^{5/4} \log\vert x \vert,
		\end{split}
	\end{equation}
	where $F(z,z_3,z_4) = \vert z-z_3 \vert^{-5/4} \vert z-z_4 \vert^{-5/4} \vert z_3-z_4 \vert^{25/24}$, the cross-ratio
	\begin{equation}
		x=\frac{(z_1-z_2)(z_3-z_4)}{(z_1-z_4)(z_3-z_2)}
	\end{equation}
	is invariant under M\"obius transformations, and $\log\vert x \vert \sim \log\vert z_1-z_2 \vert$.
	
	Combining \eqref{eq:G} with \eqref{eq:OPE} gives, asymptotically as $z_1,z_2 \to z$,
	%the contributions of all the terms in the right hand side of \eqref{eq:4-point-psi} gives}
\begin{align}
	\small
	\begin{split} \label{eq:4-point-psi-OPE-bis}
		\langle \psi(z_1)\psi(z_2)\psi(z_3)\psi(z_4) \rangle \sim \vert z_1-z_2 \vert^{-5/24} \Big[ \vert z_3-z_4 \vert^{-5/24} + \vert z_1-z_2 \vert^{5/4} F(z,z_3,z_4) \big( C_0 - C_L \log\vert x \vert \big)\Big]
		%            & \qquad \qquad - C_L \log\Big(\frac{\vert z_3-z_4 \vert}{\vert z_1-z_4 \vert \vert z_3-z_2 \vert}\Big) \Big)\Big]
	\end{split}
\end{align}
for some finite constants $C_0$ and $C_L \in (0,\infty)$.

\subsection{The OPE of the density field in the bulk and its consequences} \label{sec:bulk_analysis_OPE}

The four-point function \eqref{eq:4-point-psi-OPE-bis} is consistent with the following logarithmic OPE:
\begin{align}
	\begin{split}
		& \psi(z_1)\psi(z_2)  = \vert z_1-z_2 \vert^{-5/24} \Big[ 1 + \vert z_1-z_2 \vert^{5/4} \Big(\hat\phi(z) + \log\vert z_1-z_2 \vert \phi(z) \Big) + \ldots \Big],
	\end{split}
\end{align}
where $\hat\phi$ and $\phi$ are two distinct fields with the same scaling dimension, $2h=5/4$.
The fields $\phi$ and $\hat\phi$ are defined up to normalization, so the constants in front $\phi$ and $\hat\phi$ in the OPE have no physical meaning and can be set to one.
This choice is convenient because it leads to simple expressions for the coefficients in the two- and three-point functions below in terms of $C_0$ and $C_L$.

Inserting the OPE into $\langle\psi(z_1)\psi(z_2)\psi(z_3)\psi(z_4)\rangle$ and comparing with \eqref{eq:4-point-psi-OPE-bis} shows that
\begin{align} \label{eq:3-point-functions-hatphi}
	\langle \hat\phi(z)\psi(z_3)\psi(z_4) \rangle = \Big( C_0 - C_L \log\Big(\frac{\vert z_3-z_4 \vert}{\vert z-z_4 \vert \vert z_3-z \vert}\Big) \Big) F(z,z_3,z_4)
\end{align}
and
\begin{align} \label{eq:3-point-functions-phi}
	\langle \phi(z)\psi(z_3)\psi(z_4) \rangle = - C_L F(z,z_3,z_4) \, .
\end{align}

Using the OPE, we can formally define $\phi$ as
\begin{align} \label{eq:def-phi-bis}
	\begin{split}
		& \phi(z)  = \lim_{\varepsilon \to 0} \frac{\varepsilon^{5/24} \psi(z)\psi(z+\varepsilon u) - 1}{\varepsilon^{5/4} \vert\log\varepsilon\vert} \, ,
	\end{split}
\end{align}
where $u$ is an arbitrary vector with $\vert u \vert = 1$.

Since $\psi$ is a primary field with scaling dimension $5/48$ and two-point function
\begin{equation}
	\langle\psi(z_1)\psi(z_2)\rangle = \vert z_1-z_2 \vert^{-5/24} \, ,
\end{equation}
from the definition of $\phi$ \eqref{eq:def-phi-bis}, we see that
\begin{equation}
	\langle\phi(z)\rangle=0 \, ,
\end{equation}
as expected, but also, using \eqref{eq:3-point-functions-phi} and \eqref{eq:def-F},
\begin{align}
	\begin{split}
		& \langle\phi(z_1)\phi(z_2)\rangle = \lim_{\varepsilon \to 0} \Big\langle \frac{\varepsilon^{5/24}\psi(z_1)\psi(z_1+\varepsilon u) - 1}{\varepsilon^{5/4}\vert\log\varepsilon\vert} \, \phi(z_2) \Big\rangle \\
		& \qquad = \lim_{\varepsilon \to 0} \frac{\langle \psi(z_1)\psi(z_1+\varepsilon u)\phi(z_2)\rangle}{\varepsilon^{25/24}\vert\log\varepsilon\vert} \\
		& \qquad = \lim_{\varepsilon \to 0} \frac{-C_L \vert z_1-z_2 \vert^{-5/4} \vert z_1-z_2+\varepsilon u \vert^{-5/4}}{\vert\log\varepsilon\vert} = 0 \, .
	\end{split}
\end{align}

This calculation provides an analytic explanation for the vanishing of the two-point function of the primary field $\phi$ and shows how this is directly related to the logarithmic divergence in the four-point function \eqref{eq:4-point-psi-OPE-bis} and the associated logarithmic OPE.

We now turn to the calculation of the mixed two-point function between $\phi$ and $\hat\phi$.
Using the OPE for the density field $\psi$, for $\varepsilon$ small, we can write
\begin{align}
	\begin{split}
		& \langle \phi(z_1)\psi(z_2)\psi(z_2+\varepsilon u) \rangle = \varepsilon^{-5/24} \Big[\langle \phi(z_1) \rangle + \varepsilon^{5/4} \Big( \langle \phi(z_1)\hat\phi(z_2) \rangle + \langle \phi(z_1)\phi(z_2) \rangle \log\varepsilon \Big) + \ldots\Big] \\
		& \qquad = \varepsilon^{25/24} \langle\phi(z_1)\hat\phi(z_2)\rangle + \ldots \, .
	\end{split}
\end{align}
On the other hand, \eqref{eq:3-point-functions-phi} and \eqref{eq:def-F} imply that
\begin{align}
	\langle \phi(z_1)\psi(z_2)\psi(z_2+\varepsilon u) \rangle = -C_L \vert z_1-z_2 \vert^{-5/4} \vert z_1-z_2-\varepsilon u \vert^{-5/4} \varepsilon^{25/24} \, .
\end{align}
Comparing the last two equations and sending $\varepsilon \to 0$ gives
\begin{equation} \label{eq:2-point_function_phi-hatphi}
	\langle\phi(z_1)\hat\phi(z_2)\rangle = -C_L \vert z_1-z_2 \vert^{-5/2} \, .
\end{equation}

Similarly, we can use the OPE for the density field, combined with \eqref{eq:2-point_function_phi-hatphi}, to write
\begin{align}
	\begin{split}
		\langle \psi(z_1)\psi(z_1+\varepsilon u)\hat\phi(z_2) \rangle
		%= \varepsilon^{-5/24} \Big[\langle \hat\phi(z_1) \rangle + \varepsilon^{5/4} \Big( \langle \hat\phi(z_1)\hat\phi(z_2) \rangle + \langle \phi(z_1)\hat\phi(z_2) \rangle \log\varepsilon \Big) + \ldots\Big] \\
		= \varepsilon^{25/24} \Big( \langle\hat\phi(z_1)\hat\phi(z_2)\rangle - C_L \vert z_1-z_2 \vert^{-5/2} \log\varepsilon \Big) + \ldots \ .
	\end{split}
\end{align}
On the other hand, \eqref{eq:3-point-functions-hatphi} and \eqref{eq:def-F} imply that
\begin{align}
\begin{split}
		\langle \psi(z_1)\psi(z_1+\varepsilon u)\hat\phi(z_2) \rangle =& \Big( C_0 - C_L \log\Big(\frac{\varepsilon}{\vert z_2-z_1 \vert\vert z_1+\varepsilon u - z_2 \vert}\Big)\Big) \\
		&\quad\times\vert z_1-z_2 \vert^{-5/4} \vert z_1+\varepsilon u - z_2 \vert^{-5/4} \varepsilon^{25/24} \, .
\end{split}
\end{align}
Comparing the last two equations and sending $\varepsilon \to 0$ gives
\begin{equation}
	\langle\hat\phi(z_1)\hat\phi(z_2)\rangle = \frac{C_0 + 2 C_L \log\vert z_1-z_2 \vert}{\vert z_1-z_2 \vert^{5/2}} \, .
\end{equation}

\subsection{Analysis of the four-point function and OPE of the density field on the boundary}

As $x_1 \to x_2$, the behavior of the leading term of $\langle \psi(z)\phi_{1,3}(x_1)\phi_{1,3}(x_2)\phi_{1,3}(x_3) \rangle$ comes from the product of two half-plane two-point functions:
\begin{align}
	\begin{split}
		\langle\phi_{1,3}(x_1)\phi_{1,3}(x_2)\rangle\langle\phi_{1,3}(x_1)\phi_{1,3}(x_2)\rangle = (x_2-x_1)^{-2/3} (x_4-x_3)^{-2/3},
	\end{split}
\end{align}
%and 
%\begin{align}
%	\begin{split}
%		& \langle \psi(z)\phi_{1,3}(x_3) \rangle \sim \frac{(\Im(z))^{11/48}}{\vert z-x_3 \vert^{2/3}},
%	\end{split}
%\end{align}
where the exponents come from the value of the boundary one-arm exponent \cite{ADA99,SW01}.

The next term comes from events that require that $x_1$ and $x_2$ be the starting points of two open paths that are disjoint and separated by a closed cluster up to a certain distance, and possibly all the way to $x_3$ and $x_4$.
When $x_1$ and $x_2$ are close to each other, this produces a boundary three-arm event, with exponent $2$ \cite{ADA99,SW01}, which is responsible for the factor $(x_2-x_1)^2$.
The factor $F_H(x,x_2,x_3)$ can be obtained using the results and methods of \cite{Camia24,Camia24bis} and comes from the event, obtained in the limit as $x_1,x_2 \to x$, that $x$ is connected to $x_3$ and $x_4$ by open paths and there is a three-arm event at $x$.
A boundary three-arm event induces four interfaces, so, from a CFT perspective, it corresponds to the insertion of a four-leg operator.

To show the presence of the logarithmic terms in the 4-point function and the OPE, we can use the same arguments as in Section \ref{sec:bulk_analysis}, with disks replaced by semi-disks in the upper half-plane centered at $\frac{x_1+x_2}{2}$.
We omit the details, since they are the same as in the previous sections, and only comment on the relation between the boundary three-arm event and the boundary stress-energy tensor.

\begin{figure}
	\includegraphics[width= 0.5\textwidth]{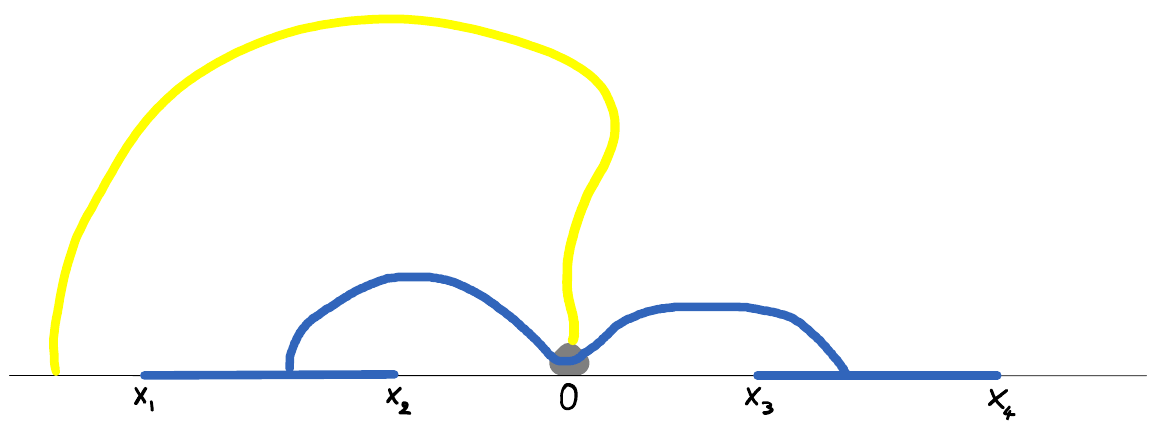}
	\caption{The event that intervals $(x_1,x_2)$ and $(x_3,x_4)$ are connected by an open (blue/dark) path in the upper half-plane, but not in the upper half-plane minus a small region adjacent to the real line near the origin $0$.}
	\label{fig:stress-energy-tensor}
\end{figure}

Let $x_1,\ldots,x_4$ be four points on the real line such that $-\infty<x_1<x_2<0<x_3<x_4<\infty$ and consider the following two events:
\begin{itemize}
	\item the interval $(x_1,x_2)$ is connected by an open path to the interval $(x_3,x_4)$ in the upper half-plane,
	\item the interval $(x_1,x_2)$ is connected by an open path to the interval $(x_3,x_4)$ in the upper half-plane minus a small region adjacent to the real line near the origin.
\end{itemize}
The difference between the probabilities of the two events is the probability of the event depicted in Figure~\ref{fig:stress-energy-tensor}, which implies the occurrence of a boundary three-arm event near the origin.
Therefore, the probability of the boundary three-arm event determines how sensitive crossing probabilities are to perturbations of the domain.
This shows that the boundary three-arm event is related to the boundary stress-energy tensor.
The fact that the boundary three-arm exponent, which is also the scaling dimension of the boundary four-leg operator, is $2$ supports this conclusion.

The four-point function \eqref{eq:4-point-psi-OPE-half-plane} is consistent with the logarithmic OPE
\begin{align} \label{eq:half-plane-logOPE-bis}
	\begin{split}
		& \phi_{1,3}(x_1)\phi_{1,3}(x_2) = (x_2-x_1)^{-2/3} \Big[ 1 + (x_2-x_1)^{2} \Big( t(x) + \log(x_2-x_1) T(x) \Big) + \ldots \Big],
	\end{split}
\end{align}
where $t$ and $T$ are two distinct fields of scaling dimension $2$.
Inserting the OPE into $\langle\phi_{1,3}(x_1)\phi_{1,3}(x_2)\phi_{1,3}(x_3)\phi_{1,3}(x_4)\rangle$ and comparing with \eqref{eq:4-point-psi-OPE-half-plane} shows that
\begin{align}
	\begin{split} \label{eq:<tphiphi>}
		\langle t(x)\phi_{1,3}(x_3)\phi_{1,3}(x_4) \rangle = \Big( C_{H,0} - C_{H,L} \log\Big(\frac{(x_4-x_3)}{(x_4-x)(x_3-x)}\Big) \Big) F_H(x,x_3,x_4)
	\end{split}
\end{align}
and
\begin{equation} \label{eq:<Tphiphi>}
    \langle T(x)\phi_{1,3}(x_3)\phi_{1,3}(x_4) \rangle = -C_{H,L} F_H(x,x_3,x_4).
\end{equation}

Using the OPE \eqref{eq:half-plane-logOPE-bis}, we can be formally define $T$ as
\begin{align} \label{eq:def-T-bis}
	\begin{split}
		& T(x)  = \lim_{\varepsilon \to 0} \frac{\varepsilon^{2/3} \phi_{1,3}(x)\phi_{1,3}(x+\varepsilon) - 1}{\varepsilon^{2} \vert\log\varepsilon\vert} \, .
	\end{split}
\end{align}

Since $\phi_{1,3}$ is a primary field with scaling dimension $1/3$ and two-point function
\begin{equation}
	\langle\phi_{1,3}(x_1)\phi_{1.3}(x_2)\rangle = \vert x_1-x_2 \vert^{-2/3} \, ,
\end{equation}
from the definition of $T$ \eqref{eq:def-T-bis}, we see that
\begin{equation}
	\langle T(x) \rangle = 0 \, ,
\end{equation}
as expected, but also, using \eqref{eq:<Tphiphi>} and \eqref{eq:def-F_H},
\begin{align}
	\begin{split}
		& \langle T(x_1) T(x_2) \rangle = \lim_{\varepsilon \to 0} \Big\langle \frac{\varepsilon^{2/3}\phi_{1,3}(x_1)\phi_{1,3}(x_1+\varepsilon) - 1}{\varepsilon^{2}\vert\log\varepsilon\vert} \, T(x_2) \Big\rangle \\
		& \qquad = \lim_{\varepsilon \to 0} \frac{\langle \phi_{1,3}(x_1)\phi_{1,3}(x_1+\varepsilon) T(x_2)\rangle}{\varepsilon^{4/3}\vert\log\varepsilon\vert} \\
		& \qquad = \lim_{\varepsilon \to 0} \frac{-C_{H,L} \vert x_1-x_2 \vert^{-2} \vert x_1-x_2+\varepsilon \vert^{-2}}{\vert\log\varepsilon\vert} = 0 \, .
	\end{split}
\end{align}
The considerations above prompt us to identify the field $T$ with the boundary stress-energy tensor.

We now turn to the calculation of the mixed two-point function between $T$ and $t$.
Using the OPE \eqref{eq:half-plane-logOPE-bis}, for $\varepsilon$ small, we can write
\begin{align}
	\begin{split}
		& \langle T(x_1) \phi_{1,3}(x_2) \phi_{1,3}(x_2 + \varepsilon u) \rangle = \varepsilon^{-2/3+2} \big( \langle T(x_1)t(x_2) \rangle + \langle T(x_1)T(x_2) \rangle \log\varepsilon \big) + \ldots \, .
	\end{split}
\end{align}
On the other hand, \eqref{eq:<Tphiphi>} and \eqref{eq:def-F_H} imply that
\begin{align}
	\langle T(x_1) \phi_{1,3}(x_2) \phi_{1,3}(x_2 + \varepsilon) \rangle = -C_L \vert x_1-x_2 \vert^{-2} \vert x_1-x_2-\varepsilon \vert^{-2} \varepsilon^{4/3} \, .
\end{align}
Comparing the last two equations and sending $\varepsilon \to 0$ gives
\begin{equation} \label{eq:<Tt>}
	\langle T(x_1) t(x_2) \rangle = -C_L \vert x_1-x_2 \vert^{-4} \, .
\end{equation}

Similarly, we can use the OPE \eqref{eq:half-plane-logOPE-bis}, combined with \eqref{eq:<Tt>}, to write
\begin{align}
	\begin{split}
		\langle \phi_{1,3}(x_1)\phi_{1,3}(x_1+\varepsilon) t(x_2) \rangle
		= \varepsilon^{4/3} \Big(\langle t(x_1)t(x_2) \rangle - C_{H,L} \vert x_1-x_2 \vert^{-4} \log\varepsilon \Big) + \ldots \ .
	\end{split}
\end{align}
On the other hand, \eqref{eq:<tphiphi>} and \eqref{eq:def-F_H} imply that
\begin{align}
\begin{split}
		\langle \phi_{1,3}(x_1)\phi_{1,3}(x_1+\varepsilon) t(x_2) \rangle =& \Big( C_{H,0} - C_{H,L} \log\Big(\frac{\varepsilon}{\vert x_2-x_1 \vert\vert x_1+\varepsilon - x_2 \vert}\Big)\Big) \\
		&\quad\times\vert x_1-x_2 \vert^{-2} \vert x_1+\varepsilon - x_2 \vert^{-2} \varepsilon^{4/3} \, .
\end{split}
\end{align}
Comparing the last two equations and sending $\varepsilon \to 0$ gives
\begin{equation}
	\langle t(x_1) t(x_2)\rangle = \frac{C_{H,0} + 2 C_{H,L} \log\vert x_1-x_2 \vert}{\vert x_1-x_2 \vert^{4}}.
\end{equation}

We conclude that $t$ is the logarithmic partner to the stress-energy tensor $T$.

\subsection{Analysis of $\langle \phi_{1,4}(x_1) \phi_{1,4}(x_2) \phi_{1,2}(x_3) \phi_{1,2}(x_4) \rangle$} \label{sec:analysis_mixed}
In the discussion below, we focus on critical percolation in the upper half-plane, with the real line as the boundary.
In the figures, we draw open paths in blue (darker lines), closed paths in yellow (lighter lines) and the interfaces between open and closed clusters in green (light, thin, wiggly lines).

\begin{figure}
	\includegraphics[width= 0.7\textwidth]{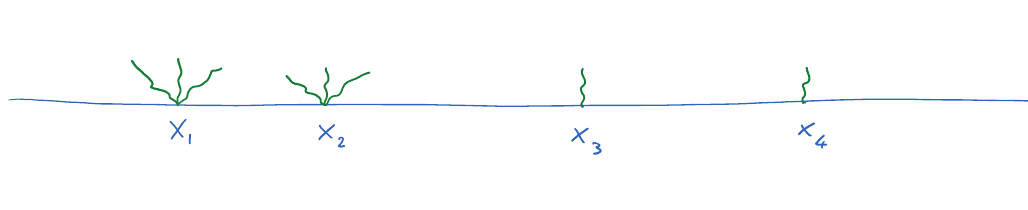}
	\caption{The $4$-point function $\langle \phi_{1,4}(x_1) \phi_{1,4}(x_2) \phi_{1,2}(x_3) \phi_{1,2}(x_4) \rangle$ corresponds to inserting three interfaces at $x_1$ and at $x_2$ and one interface at $x_3$ and at $x_4$.}
	\label{fig:SLE_insertion}
\end{figure}

\begin{figure}
	\includegraphics[width= 0.5\textwidth]{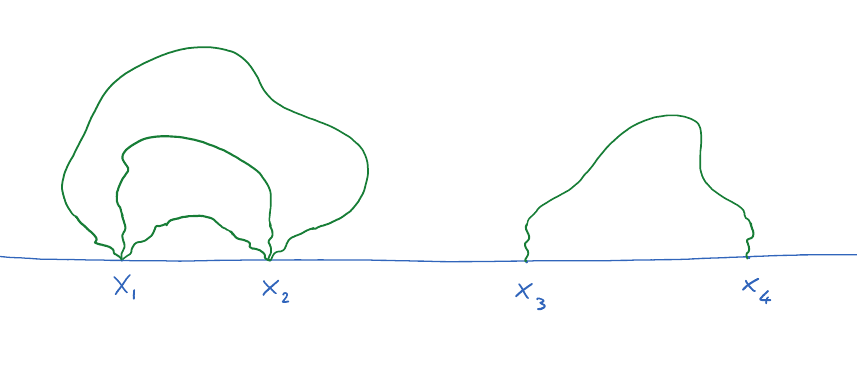}
	\caption{Matching of the interfaces that does not connect $x_1$ and $x_2$ to $x_3$ and $x_4$.}
	\label{fig:power-law_matching}
\end{figure}

\begin{figure}
	\includegraphics[width= 0.5\textwidth]{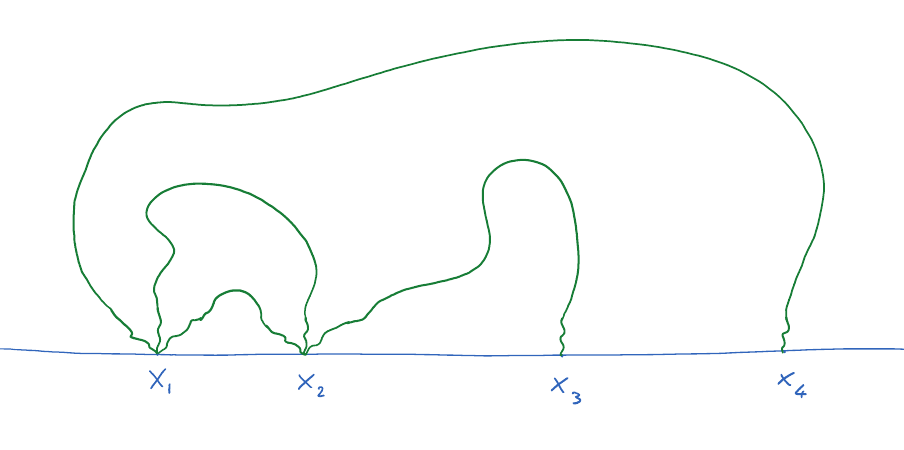}
	\caption{Matching of the interfaces that connects $x_1$ to $x_4$ and $x_2$ to $x_3$.}
	\label{fig:nontrivial_matching}
\end{figure}

As already mentioned, the four-point function $\langle \phi_{1,4}(x_1) \phi_{1,4}(x_2) \phi_{1,2}(x_3) \phi_{1,2}(x_4) \rangle$ corresponds to the insertion of three interfaces at $x_1$ and $x_2$ and one interface at $x_3$ and $x_4$, as represented in Fig.~\ref{fig:SLE_insertion}.
When connecting the interfaces, we need to consider two situations, one in which the interfaces from $x_1$ and $x_2$ are not connected to those from $x_3$ and $x_4$, as depicted in Figure~\ref{fig:power-law_matching}, and one in which the interfaces from $x_1$ and $x_2$ are connected to those from $x_3$ and $x_4$, as in Figure~\ref{fig:nontrivial_matching}.
With this in mind, given that $\phi_{1,4}$ has scaling dimension $1$, we can write
\begin{equation} \label{eq:4-point-func}
	\langle \phi_{1,4}(x_1) \phi_{1,4}(x_2) \phi_{1,2}(x_3) \phi_{1,2}(x_4) \rangle = \vert x_1-x_2 \vert^{-2} \big(F_d(x)+F_c(x)\big),
\end{equation}
where the functions $F_d$ and $F_c$, coming from the ``disconnected'' diagram in Fig.~\ref{fig:power-law_matching} and the ``connected'' diagram in Fig.~\ref{fig:nontrivial_matching}, respectively, depend only on the cross-ratio
\begin{equation} \label{cross-ratio}
	x=\frac{(x_4-x_2)(x_3-x_1)}{(x_3-x_2)(x_4-x_1)},
\end{equation}
which is invariant under conformal transformations.

Observing that the ``probability'' of the event depicted in Fig.~\ref{fig:power-law_matching} satisfies
%\footnote{We emphasize that the function $P$ should be interpreted as the continuum limit of the renormalized lattice probability (the renormalization factor is the inverse of the square of the lattice spacing, see Theorem \ref{theorem} below) and can therefore be larger than $1$.}
\begin{align} \label{eq:Fig.2.2}
	\lim_{x_4 \to x_3} P(\text{Fig.}~\ref{fig:power-law_matching}) = \langle \phi_{1,4}(x_1) \phi_{1,4}(x_2) \rangle = \vert x_1-x_2 \vert^{-2},
\end{align}
we conclude that $F_d(x)=1+f_1(x)$ with $\lim_{x_4 \to x_3} f_1(x)=0$.
As $x_4 \to x_3$, the leading term in $F_d$ comes from the probability that the interface from $x_3$ to $x_4$ makes a small excursion from the boundary, while the next term, $f_1$, contains the probability that it makes a large excursion, corresponding to a macroscopic cluster anchored to the interval $(x_3,x_4)$, or equivalently, to a boundary one-arm event.
Since the boundary one-arm exponent is $1/3$ \cite{ADA99,SW01}, we see that $f_1(x) \sim (x_4-x_3)^{1/3}$ as $x_4 \to x_3$.

When $x_1 \to x_2$, $f_1(x) \sim (x_2-x_1)^{1/3}+O\big((x_2-x_1)^2\big)$,
%\begin{align}
%    & P(\text{Fig.}~\ref{fig:power-law_matching}) \sim \vert x_1-x_2 \vert^{-2}\Big(1 + \text{const} (x_2-x_1)^{1/3} (x_3-x_1)^{-1/3} (x_4-x_1)^{-1/3} (x_4-x_3)^{1/3} \\
%    & \qquad \qquad + O\big((x_2-x_1)^{2}\big)\Big),
%\end{align}
where the first term comes from the event that the outermost interface from $x_1$ to $x_2$ makes a large excursion, producing a boundary one-arm event, and the second one comes from the event that both the outermost and the middle interface make large excursions, producing a boundary three-arm event, with exponent $2$ \cite{ADA99,SW01}.
%This means that $f_1(x) \sim (x_2-x_1)^{1/3}$ as $x_1 \to x_2$.

\begin{figure}
	\includegraphics[width= 0.5\textwidth]{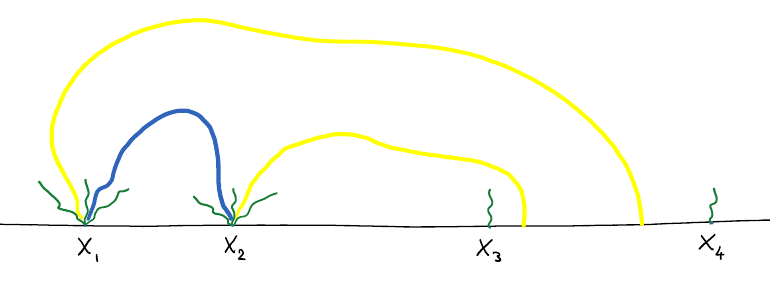}
	\caption{The matching of Fig.~\ref{fig:nontrivial_matching} in terms of paths with two disjoint closed (yellow/light) paths landing in the interval $(x_3,x_4)$.}
	\label{fig:disjoint_paths_red}
\end{figure}

\begin{figure}
	\includegraphics[width= 0.5\textwidth]{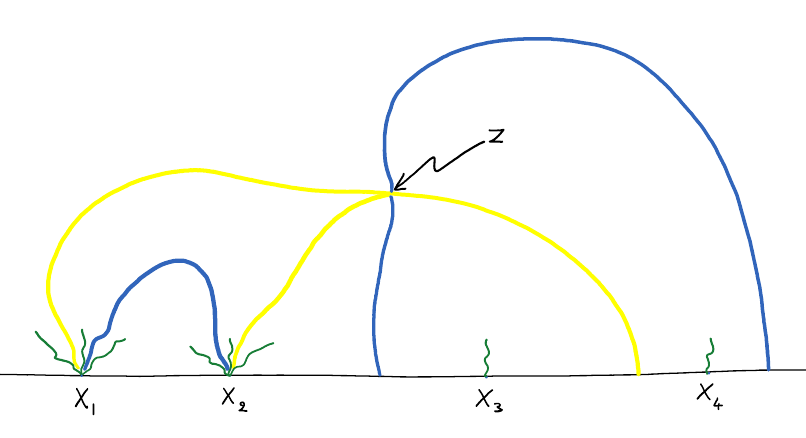}
	\caption{The matching of Fig.~\ref{fig:nontrivial_matching} in terms of paths with a single closed (yellow/light) path landing in the interval $(x_3,x_4)$.}
	\label{fig:pinched_path}
\end{figure}

The matching of Fig.~\ref{fig:nontrivial_matching} corresponds, in terms of clusters or paths, to one of the two situations depicted in Figs.~\ref{fig:disjoint_paths_red} and \ref{fig:pinched_path}.
In the first case, there are two disjoint yellow paths from $x_1$ and $x_2$ landing in the interval $(x_3,x_4)$.
We postpone the analysis of this case to the next section.
In the second case, the two yellow paths starting at $x_1$ and $x_2$ are forced to join by two blue clusters pinching the yellow cluster at some point $z$ in the upper half-plane.
%anchored to $x_1$ and $x_2$, producing a pivotal point at $z_5$, corresponding to a five-arm event in the bulk.

If we send $x_4 \to x_3$, we see that the contributions coming from the configurations depicted in Figs.~\ref{fig:disjoint_paths_red} and \ref{fig:pinched_path} must go to zero like $\vert x_3-x_4 \vert$ and $\vert x_3-x_4 \vert^{1/3}$, respectively, because the boundary two-arm and one-arm exponents are, respectively, $1$ and $1/3$ \cite{ADA99,SW01}.
Because of this, and considering that, as $x_4 \to x_3$, the configuration depicted in Fig.~\ref{fig:pinched_path} becomes one with two arms, one blue and one yellow, at $x_1$ and $x_2$ and one yellow arm at $x_3$, we see that
%The boundary one-arm exponent is $1/3$ \cite{ADA99,SW01} and the presence of a boundary one-arm event corresponds to the insertion of the boundary operator $\phi_{1,3}$ with conformal dimension $1/3$.
%Therefore, we see that
\begin{align} \label{eq:limit_pinched_path}
	\begin{split}
		& \lim_{x_4 \to x_3} \frac{P(\text{Fig.~\ref{fig:nontrivial_matching}})}{{\vert x_3-x_4 \vert^{1/3}}} = \lim_{x_4 \to x_3} \frac{P(\text{Fig.~\ref{fig:pinched_path}})}{{\vert x_3-x_4 \vert^{1/3}}} \sim \langle \phi_{1,4}(x_1) \phi_{1,4}(x_2) \phi_{1,3}(x_3) \rangle \\
		& \qquad \sim \vert x_1-x_2 \vert^{-5/3} \vert x_1-x_3 \vert^{-1/3} \vert x_2-x_3 \vert^{-1/3}.
	\end{split}
\end{align}
%where $C_{\phi_{1,4},\phi_{1,4},\phi_{1,3}}$ is one of the structure constants of the theory.

When $x_1$ is close to $x_2$, the blue path connecting $x_1$ and $x_2$ tends to be small and the pinch point $z$ tends to be close to the interval $(x_1,x_2)$, producing a (yellow) boundary one-arm event. The next term corresponds to a macroscopic blue cluster anchored to $x_1$ and $x_2$ and producing a boundary three-arm event. Since the boundary one-arm and three-arm exponents are, respectively, $1/3$ and $2$ \cite{ADA99,SW01}, we can write $P(\text{Fig.~\ref{fig:pinched_path}})=(x_2-x_1)^2f_2(x)$, where $f_2$ depends only on the cross-ratio $x$ and
\begin{align}
	f_2(x) \sim (x_2-x_1)^{1/3} + O\big((x_2-x_1)^{2}\big) \;\;\; \text{ as } x_1 \to x_2.
\end{align}

%\begin{align}
%& P(\text{Fig.~\ref{fig:pinched_path}}) \sim (x_2-x_1)^{-2}\big[C_{\phi_{1,3},\phi_{1,2},\phi_{1,2}}(x_3-x_2)^{-1/3}(x_4-x_2)^{-1/3}(x_4-x_3)^{1/3}(x_2-x_1)^{1/3} \\
%& \qquad \qquad \qquad \qquad \qquad \qquad + O\big((x_2-%x_1)^{2}\big)\big],
%& \qquad=(x_2-x_1)^{-5/3}+O(1)
%\end{align}
%\textcolor{blue}{Based on these observations, we can write the contribution coming from Fig.~\ref{fig:pinched_path} as $(x_2-x_1)^{-2}f_2(x)$, where
	%\begin{align}
	%f_2(x) = C_{\phi_{1,3},\phi_{1,2},\phi_{1,2}}(x_3-x_2)^{-1/3}(x_4-x_2)^{-1/3}(x_4-x_3)^{1/3}(x_2-x_1)^{1/3} + O\big((x_2-x_1)^{2}\big)    
	%\end{align}
	%}

We note that we have discussed so far
%namely \eqref{eq:limit_power-law_matching}, \eqref{eq:limit_pinched_path} and $\lim_{x_4 \to x_3} P(\text{Fig.~\ref{fig:disjoint_paths_red}})=0$,
is consistent with the OPE \eqref{eq:OPE_phi12}.
Indeed, given that the conformal dimensions of $\phi_{1,2}$ and $\phi_{1,3}$ are $h_{1,2}=0$ and $h_{1,3}=1/3$, respectively, \eqref{eq:OPE_phi12} leads to
\begin{align} 
	\small
	\begin{split} \label{eq:OPE_applied}
		& \langle \phi_{1,4}(x_1) \phi_{1,4}(x_2) \phi_{1,2}(x_3) \phi_{1,2}(x_4) \rangle \\
		& \quad = \langle \phi_{1,2}(x_3) \phi_{1,2}(x_4) \rangle \Big( \langle \phi_{1,4}(x_1) \phi_{1,4}(x_2) \rangle + \langle \phi_{1,4}(x_1) \phi_{1,4}(x_2) \phi_{1,3}(x_3) \rangle \vert x_3-x_4 \vert^{h_{1,3}} + O\big(\vert x_3-x_4 \vert^{h_{1,3}}\big) \Big) \\
		& \quad = \vert x_1-x_2 \vert^{-2} + C_{\phi_{1,4},\phi_{1,4},\phi_{1,3}} \vert x_1-x_2 \vert^{-5/3} \vert x_1-x_3 \vert^{-1/3} \vert x_2-x_3 \vert^{-1/3} \vert x_3-x_4 \vert^{1/3} + O\big(\vert x_3-x_4 \vert^{1/3}\big) \\
		& \quad = \vert x_1-x_2 \vert^{-2} \Big( 1 + C_{\phi_{1,4},\phi_{1,4},\phi_{1,3}} \vert x_1-x_2 \vert^{1/3} \vert x_1-x_3 \vert^{-1/3} \vert x_2-x_3 \vert^{-1/3} \vert x_3-x_4 \vert^{1/3} + O\big(\vert x_3-x_4 \vert^{1/3}\big) \Big),
	\end{split}
\end{align}
where $C_{\phi_{1,4},\phi_{1,4},\phi_{1,3}}$ is one of the structure constants of the theory.

%\begin{align} 
%\begin{split} \label{eq:lim_x4->x3}
%    & \lim_{x_4 \to x_3} \frac{\langle \phi_{1,4}(x_1) \phi_{1,4}(x_2) \phi_{1,2}(x_3) \phi_{1,2}(x_4) \rangle}{\langle \phi_{1,2}(x_3) \phi_{1,2}(x_4) \rangle} = \langle \phi_{1,4}(x_1) \phi_{1,4}(x_2) \rangle + \langle \phi_{1,4}(x_1) \phi_{1,4}(x_2) \phi_{1,3}(x_3) \rangle \\
%    & \qquad = \vert x_1-x_2 \vert^{-2} + C_{\phi_{1,4},\phi_{1,4},\phi_{1,3}} \vert x_1-x_2 \vert^{-5/3} \vert x_1-x_3 \vert^{-1/3} \vert x_2-x_3 \vert^{-1/3}. %\\
%    & \qquad = \vert x_1-x_2 \vert^{-2} \Big( 1 + C_{\phi_{1,4},\phi_{1,4},\phi_{1,3}} \vert x_1-x_2 \vert^{1/3} \vert x_1-x_3 \vert^{-1/3} \vert x_2-x_3 \vert^{-1/3}\Big).
%\end{split}
%\end{align}
%This means that the term corresponding to Fig.~\ref{fig:power-law_matching} tends to $\langle \phi_{1,4} \phi_{1,4} \rangle = \vert x_1-x_2 \vert^{-2}$ as $x_4 \to x_3$.

%\section{Logarithmic term}
%We conclude our study of the four-point function $\langle \phi_{1,4}(x_1) \phi_{1,4}(x_2) \phi_{1,2}(x_3) \phi_{1,2}(x_4) \rangle$ by analyzing
We now analyze in greater detail the term $P(\text{Fig.}~\ref{fig:disjoint_paths_red})$, which vanishes as $\vert x_3-x_4 \vert$ when $x_4 \to x_3$, as we argued above,
and therefore does not appear in \eqref{eq:OPE_applied}.
%and is therefore not visible in \eqref{eq:lim_x4->x3}.
%Our analysis of this term uses the so-called \emph{color-switching trick}, which can only be performed for (Bernoulli) percolation, therefore it does not extend to other Fortuin-Kasteleyn (FK) percolation models.

\begin{figure}
	\includegraphics[width= 0.5\textwidth]{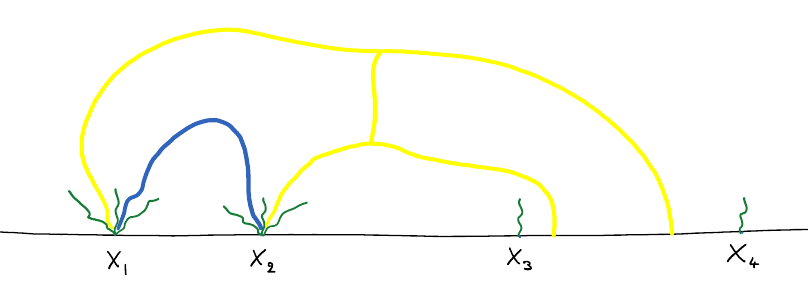}
	\caption{The situation depicted in Fig.~\ref{fig:disjoint_paths_red} with the two closed (yellow/light) paths belonging to the same cluster. In this case, the two closed (yellow/light) paths must be connected by a closed (yellow/light) path.}
	\label{fig:same_cluster}
\end{figure}

\begin{figure}
	\includegraphics[width= 0.5\textwidth]{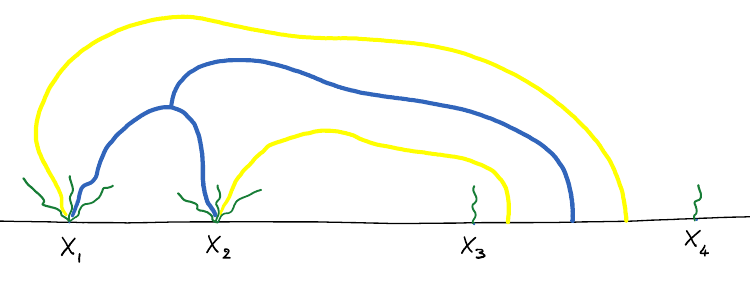}
	\caption{The situation depicted in Fig.~\ref{fig:disjoint_paths_red} with the two closed (yellow/light) paths belonging to different clusters. In this case, the two closed (yellow/light) paths must be separated by an open (blue/dark) cluster.}
	\label{fig:three_paths}
\end{figure}

Because of topological considerations and the duality between open (blue) and closed (yellow) paths, Fig.~\ref{fig:disjoint_paths_red} corresponds to two types of situations: one in which the two yellow paths belong to the same cluster, and are therefore connected by a yellow path (Fig.~\ref{fig:same_cluster}), and one in which they belong to different clusters, which implies the presence of a blue path separating them (Fig.~\ref{fig:three_paths}).

In the latter situation, we have three arms of alternating color (yellow, blue, yellow) landing in the interval $(x_3,x_4)$.
This contribution must scale like the other terms in the four-point function, that is,
\begin{equation} \label{eq:three-paths}
	P(\text{Fig.}~\ref{fig:three_paths}) = \vert x_1-x_2 \vert^{-2} \, f_3(x),
\end{equation}
with a function $f_3$ that depends only on the cross-ratio $x$,
%\begin{equation} \label{cross-ratio}
%    x=\frac{(x_4-x_2)(x_3-x_1)}{(x_3-x_2)(x_4-x_1)},
%\end{equation}
which is invariant under conformal maps.
If we send $x_4 \to x_3$, Fig.~\ref{fig:three_paths} produces a boundary three-arm event at $x_3$, corresponding to the insertion of a boundary four-leg operator.
Since the boundary three-arm exponent is $2$ \cite{ADA99,SW01}, we conclude that
\begin{equation} \label{eq:f}
	f_3(x) \sim \vert x_3-x_4 \vert^{2} \text{ as } x_4 \to x_3.
\end{equation}
Sending, instead, $x_1 \to x_2$ produces a boundary three-arm event from the interval $(x_1,x_2)$, giving
\begin{equation} \label{eq:f2}
	f_3(x) \sim \vert x_1-x_2 \vert^{2} \text{ as } x_1 \to x_2.
\end{equation} 


\begin{figure}
	\includegraphics[width= 0.5\textwidth]{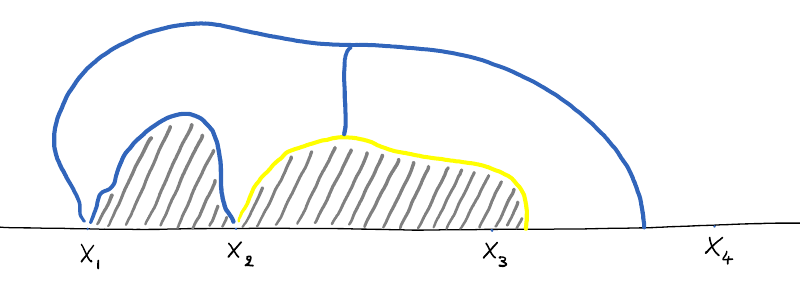}
	\caption{Starting from the situation in Fig.~\ref{fig:same_cluster}, an exploration process and the switching of colors outside the explored regions lead to the situation depicted here.}
	\label{fig:color-switch}
\end{figure}

We now turn to the situation depicted in Fig.~\ref{fig:same_cluster}.
For every configuration corresponding to that situation, 
%depicted in  Fig.~\ref{fig:disjoint_paths_red},
there is a configuration with the same probability with the top yellow path replaced by a blue path. This can be seen by an exploration of the regions below the lowest blue path between $x_1$ and $x_2$ and below the lowest yellow path starting at $x_2$ and landing in the interval $(x_3,x_4)$ (the shaded regions in Fig.~\ref{fig:color-switch}).
It is a standard result of percolation theory that one can perform such explorations without gathering any information about the regions above the lowest path.
Outside the explored regions, there is a one-to-one correspondence between a given configuration and one with all colors switched (see Fig.~\ref{fig:color-switch}).
%configurations with a red path to the interval $(x_3,x_4)$, as in Fig.~\ref{fig:same_cluster}, and configurations with a black path to the interval $(x_3,x_4)$, as in Fig.~\ref{fig:color-switch}.
%To see this, just take any configuration of the first type and change all red clusters to black and vice versa outside the explored regions (see Fig.~\ref{fig:color-switch}).
Moreover, because of the symmetry between blue and yellow, switching all colors in a given region does not change the probability of the configuration.
%this color-switching procedure preserves the probability $P$; that is, the final configuration has the same probability as the initial one.

\begin{figure}
	\includegraphics[width= 0.5\textwidth]{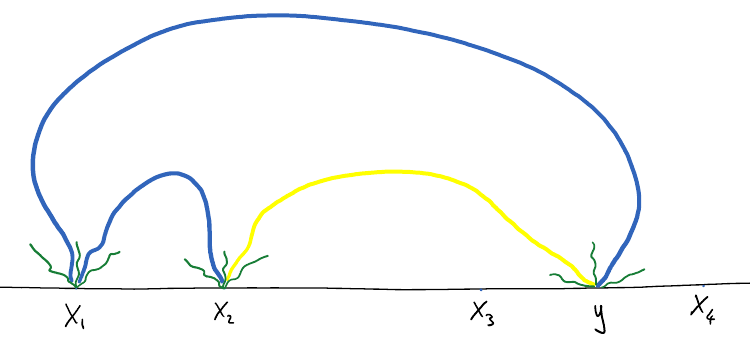}
	\caption{In Fig.~\ref{fig:color-switch}, the closed (yellow/light) cluster and the open (blue/dark) cluster landing in the interval $(x_3,x_4)$ are adjacent. Therefore, there must be a boundary two-arm event somewhere in the interval $(x_3,x_4)$.}
	\label{fig:3-point_function}
\end{figure}

The blue and yellow clusters landing in the interval $(x_3,x_4)$ in Fig.~\ref{fig:color-switch} are adjacent.
Therefore, they must meet on the interval and produce a boundary two-arm event at a unique position $y \in (x_3,x_4)$, as depicted in Fig.~\ref{fig:3-point_function}.
The conformal invariance of the scaling limit of critical percolation \cite{CamiaNewmanPercolationFull} and the scaling dimension of the boundary two-arm event imply that 
$P(\text{Fig.}~\ref{fig:3-point_function}) = C \, \vert x_1-x_2 \vert^{-1} \vert x_1-y \vert^{-1} \vert x_2-y \vert^{-1}$ for some constant $C \in (0,\infty)$.
Alternatively, since a boundary two-arm event corresponds to the insertion of a boundary three-leg operator, one can see that $P(\text{Fig.}~\ref{fig:3-point_function})$ must scale like the three-point function $\langle \phi_{1,4}(x_1) \phi_{1,4}(x_2) \phi_{1,4}(y) \rangle$.

Therefore, we obtain
\begin{equation} \label{eq:log-term}
	P(\text{Fig.}~\ref{fig:same_cluster}) = C \, \vert x_1-x_2 \vert^{-1} \int_{x_3}^{x_4} \vert x_1-y \vert^{-1} \vert x_2-y \vert^{-1} \, dy = C \, \vert x_1-x_2 \vert^{-2} \log \frac{(x_4-x_2)(x_3-x_1)}{(x_3-x_2)(x_4-x_1)},
\end{equation}
which shows the presence of a logarithmic term in the four-point function
\[\langle \phi_{1,4}(x_1) \phi_{1,4}(x_2) \phi_{1,2}(x_3) \phi_{1,2}(x_4) \rangle.\]
Indeed, summing up the results of this section and the previous one, %namely %\eqref{eq:lim_x4->x3}
%\eqref{eq:4-point-func}, \eqref{eq:OPE_applied}, \eqref{eq:three-paths}, \eqref{eq:f} and \eqref{eq:log-term},
we can write
\begin{align}
	\langle \phi_{1,4}(x_1) \phi_{1,4}(x_2) \phi_{1,2}(x_3) \phi_{1,2}(x_4) \rangle = \vert x_1-x_2 \vert^{-2} \big( 1 + f_0(x) + C \log x \big),
\end{align}
where $f_0(x) = f_1(x) + f_2(x) + f_3(x)$ and the asymptotic behavior of $f_1(x),f_2(x)$ and $f_3(x)$ for $x_4 \to x_3$ and $x_1 \to x_2$ was discussed earlier.

When $x_3 \to x_2$, one can see from Fig.~\ref{fig:power-law_matching} that the disconnected diagram leads to a three-arm event at $x_2$, so that its contribution tends to zero.
The connected diagram, instead, leads to a one-arm event at $x_2$, whose probability goes to zero like $(x_3-x_2)^{1/3}$.
However, the ``probability'' $P$ is normalized using the two-arm exponent, equal to $1$, at $x_2$.
Therefore, $\hat{f}(x) \sim P(\text{Fig.}~\ref{fig:power-law_matching}) \sim (x_3-x_2)^{-1+1/3} = (x_3-x_2)^{-2/3}$ as $x_3-x_2 \to 0$.

\section{Conclusions}

We have provided several examples of correlation functions for critical percolation involving a logarithmic singularity.
One of them is the four-point function of the density (spin) field $\psi$ in the bulk, which has a logarithmic divergence as two of the points collide.
We have shown that the logarithmic divergence appears as a consequence of scale invariance combined with independence, which is a manifestation of the fact that percolation has central charge $c=0$.

More precisely, as $z_1,z_2 \to z$, the four-point function $\langle\psi(z_1)\psi(z_2)\psi(z_3)\psi(z_4)\rangle$ contains terms that correspond to the probabilities of events of the following type:
\begin{itemize}
	\item there are macroscopic open paths starting at $z_1$ and $z_2$ and
	\item there is an $r \geq 2\vert z_1-z_2 \vert$ such that, within the disk $B(r)$ of radius $r$ centered at $z$, the two paths are disjoint and separated by a closed cluster (so that, within $B(r)$, the event is equivalent to the insertion of a four-leg operator) and
	\item the two open paths ``hook up'' within the annulus $B(2r) \setminus B(r)$.
\end{itemize}
Due to scale invariance, the probabilities of events of this type are of the same order for different values of $r$.
The logarithm in \eqref{eq:4-point-psi-OPE}, and consequently in the OPE \eqref{eq:logOPE}, corresponds to the number of annuli $B(2r) \setminus B(r)$ one can insert in the space between $z_1,z_2$ and $z_3,z_4$, starting from the disk of radius $r=2\vert z_1-z_2 \vert$ and doubling the radius at each step, which is of order $\log\frac{1}{\vert z_1-z_2 \vert}$.
%Independence, which manifests itself in the value of the central charge, $c=0$, is crucial in the analysis leading to \eqref{eq:summation}.
%Scale invariance implies that these terms are of the same order over a range of scales of order $\log\frac{1}{\vert z_1-z_2 \vert}$, leading to a logarithmic divergence as $\vert z_1-z_2 \vert \to 0$.
%We believe that this mechanism is quite general and can explain other logarithmic singularities, at least in the context of critical percolation~\cite{CF24}.

The same explanation applies to the logarithmic divergence found in our second example, the four-point function between density fields on the boundary.

A similar mechanism is at work in the case of our third example, the correlation between two boundary three-leg operators and two boundary one-leg operators.
One contribution to the four-point function $\langle \phi_{1,4}(x_1) \phi_{1,4}(x_2) \phi_{1,2}(x_3) \phi_{1,2}(x_4) \rangle$ comes from the ``sum'' over $y$ between $x_3$ and $x_4$ of the probabilities of events of the type depicted in Fig.~\ref{fig:3-point_function}, which takes the form of an integral over $(x_3,x_4)$ of a three-point function.
The integral can be easily computed and is proportional to the logarithm of the cross-ratio between $x_1, x_2, x_3, x_4$.

The asymptotic behavior of the first two four-point functions discussed above is consistent with logarithmic OPEs containing pairs of new fields with the same scaling dimension.
The correlation functions of the new fields show that they are logarithmic partners \cite{Gur93,CR13}.

In the bulk case, one can identify the new field $\phi$ with (a multiple of) the energy field (discussed, for instance, in Section~4.2 of~\cite{Car13}).
Moreover, the analysis leading to \eqref{eq:4-point-psi-OPE} shows that $P(z_1 \leftrightarrow z_2, z_3 \leftrightarrow z_4) = P(z_1 \leftrightarrow z_2 \leftrightarrow z_3 \leftrightarrow z_4) + P(z_1 \leftrightarrow z_2 \not\leftrightarrow z_3 \leftrightarrow z_4)$ contributes to both subleading terms, while $P(z_1 \leftrightarrow z_3 \not\leftrightarrow z_2 \leftrightarrow z_4)$ and $P(z_1 \leftrightarrow z_4 \not\leftrightarrow z_2 \leftrightarrow z_3)$, which correspond to the insertion of two distinct open clusters,
% four-leg operator (equivalently, the occurrence of a four-arm event or the insertion of two distinct open clusters),
contribute only to the subleading term without the logarithm.
In other words, while the term with the logarithm receives contributions only from $P(z_1 \leftrightarrow z_2, z_3 \leftrightarrow z_4)$, the subleading term without the logarithm receives contributions from both $P(z_1 \leftrightarrow z_2, z_3 \leftrightarrow z_4)$ and $P(z_1 \leftrightarrow z_3 \not\leftrightarrow z_2 \leftrightarrow z_4)+P(z_1 \leftrightarrow z_4 \not\leftrightarrow z_2 \leftrightarrow z_3)$.
%This, combined with \eqref{eq:logOPE} and \eqref{eq:2-point-functions}, shows
This means that we can think of $\hat\phi$ as a mixture of $\phi$ and a field that creates two separate open clusters.\footnote{We thank Rongvoram Nivesvivat for this observation.}
This is consistent with the conclusions of \cite{VJS12}, reached by analyzing the behavior of correlation functions in the $Q$-state Potts model with $Q \neq 1$ and taking the (formal) limit $Q \to 1$, where the logarithmic term is also attributed to the mixing of the energy field with a field that creates two propagating clusters.

In the boundary case, one of the new fields emerging from the OPE of two density fields placed on the boundary can be identified with (a multiple of) the stress-energy tensor and the second field with its logarithmic partner.
The latter can be interpreted as a mixture of the stress-energy tensor and a field that creates two separate open clusters anchored to a point on the boundary.
The appearance of a logarithmic partner to the stress-energy tensor is consistent with Gurarie's proposed solution of the ``$c=0$ catastrophe'' \cite{Gur93}.
%second field with the same scaling dimension is consistent with Gurarie's proposal that the presence of a logarithmic partner to the stress-energy tensor may be necessary to avoid the ``$c=0$ catastrophe'' \cite{Gur93}.
Also in this case, our conclusions are consistent with previous results, obtained with different methods.
%on the boundary stress-energy tensor, obtained by analyzing the structure of indecomposable Virasoro modules.
For instance, the appearance of a logarithmic singularity was predicted in \cite{GV18} solving a differential equation.
Moreover, the correlation functions of the stress-energy tensor and its logarithmic partner can be derived by analyzing the structure of indecomposable Virasoro modules.
In particular, our equations \eqref{eq:2-point-functions-boundary} correspond to equations (1.12) of \cite{VJS11} and, upon identifying $t$ with a multiple of $\phi_{1,5}$ (possibly mixed with $T$), with equations (3.14) and (3.15) of~\cite{mathieu2007percolation}.

The findings of Section \ref{sec:mixed} are also consistent with previous results, in particular with the fusion rules $\phi_{1,2} \times \phi_{1,2} = \phi_{1,3} + \ldots$ and $\phi_{1,4} \times \phi_{1,4} = \phi_{1,3} + \ldots$ , derived in~\cite{mathieu2007percolation} using algebraic methods.

Since our approach is purely probabilistic, involves only percolation observables, and is based on rigorous results for the scaling limit of percolation \cite{Camia24,Camia24bis,CF24}, our analysis not only confirms the logarithmic nature of the percolation CFT, but can be seen as a further test of the assumptions behind the algebraic and bootstrap techniques.
%approach and of the assumption that the $Q \to 1$ limit of Potts correlation functions reproduces percolation correlations.}

It would be interesting to explore if our methods can be applied to higher dimensions and to other models, such as the FK random cluster model and the $O(n)$ model, and to combine them with other CFT tools.
Indeed, after reading a draft version of this paper, 
%We conclude this discussion by mentioning that, as explained to us by Rongvoram Nivesvivat\footnote{Private communication. We thank Rongvoram Nivesvivat for sharing his insight with us and pointing us to the relevant literature.}
using the results for the Potts model obtained in~\cite{NRJ23} via the algebraic approach mentioned above, combined with numerical conformal bootstrap methods (see~\cite{JS19,NRJ23,PRS19,HJS20,NR21}), and taking the $Q \to 1$ limit of the four-point function of the $Q$-state Potts model spin field, Rongvoram Nivesvivat\footnote{Private communication. We thank Rongvoram Nivesvivat for sharing his insight with us and pointing us to the relevant literature.} was able to reproduce the asymptotic expansion \eqref{eq:4-point-psi-OPE} and to calculate (numerically) the constants $C_0$ and $C_L$ that appear in the equation.
%Considering that our analysis relies only on percolation observables and techniques, this can be seen as a further test of the assumptions behind the algebraic and bootstrap techniques.

%our derivation of \eqref{eq:4-point-psi-OPE} can be considered both as an independent verification of its validity and as a further test of the assumptions behind the algebraic and bootstrap techniques.
%In view of this observation, given that our analysis leads to \eqref{eq:4-point-psi-OPE} using only percolation observables and techniques, without any additional assumptions, our results provide further support for the validity of the assumptions behind the algebraic and bootstrap approaches and for the fact that the $Q \to 1$ limit of Potts correlations correctly reproduces percolation correlations.

%While it is interesting to have an alternative derivation of some of our results, a main message of this paper is that it is possible to show the appearance of logarithmic terms in percolation four-point functions and OPEs using only percolation observables and techniques, without any additional assumptions.
%Moreover, by doing so, one gains new insight on the mechanism that leads to logarithmic terms in critical percolation and, by expressing connectivities in terms of field correlators, on the geometric meaning of the fields involved.

\acknowledgments

The authors benefited from conversations with and comments from Gesualdo Delfino, Matthew Kleban, Peter Kleban, Kalle Kyt\"ol\"a, Chuck Newman, Sylvain Ribault and especially Rongvoram Nivesvivat.
Yu Feng thanks NYUAD for its hospitality during a visit in the fall of 2023, when this project was started.
The visit was partially supported by the Short-Term Visiting Fund for Doctoral Students of Tsinghua University.

%\paragraph{Note added.} This is also a good position for notes added
%after the paper has been written.

% Bibliography

%% [A] Recommended: using JHEP.bst file
%% \bibliographystyle{JHEP}
%% \bibliography{biblio.bib}

%% or
%% [B] Manual formatting (see below)
%% (i) We suggest to always provide author, title and journal data or doi:
%% in short all the informations that clearly identify a document.
%% (ii) please avoid comments such as "For a review'', "For some examples",
%% "and references therein" or move them in the text. In general, please leave only references in the bibliography and move all
%% accessory text in footnotes.
%% (iii) Also, please have only one work for each \bibitem.

\bibliographystyle{JHEP}
\bibliography{bibliography}
%\begin{thebibliography}{99}
%
%\bibitem{a}
%Author,
%\emph{Title},
%\emph{J. Abbrev.} {\bf vol} (year) pg.
%
%\bibitem{b}
%Author,
%\emph{Title},
%arxiv:1234.5678.
%
%\bibitem{c}
%Author,
%\emph{Title},
%Publisher (year).
%
%\end{thebibliography}
\end{document}